\documentclass[printer]{aa} 

\usepackage{graphicx}
\usepackage{txfonts}
\usepackage{color}
\usepackage{hyperref}

\def\fdeg{\hbox{$^\circ$}}
\def\degr{^\circ}

\newcommand{\BP}{G_{BP}}
\newcommand{\RP}{G_{RP}}
\newcommand{\GKO}{(G-K)_0}
\newcommand{\AO}{A_0}
\newcommand{\Teff}{{\mathrm{T_{eff}}}}
\newcommand{\logg}{\log g}

\begin{document}

   \title{Gaia-2MASS 3D maps of Galactic interstellar dust within 3 kpc}


  \author{R. Lallement
\inst{1}
\and
C. Babusiaux\inst{2,1}
\and
J.L. Vergely\inst{3}
\and 
D. Katz\inst{1}
\and
F. Arenou\inst{1}
\and
B. Valette\inst{4}
\and
C. Hottier\inst{1}
\and
L. Capitanio\inst{1}
}

\institute{GEPI, Observatoire de Paris, PSL University, CNRS,  5 Place Jules Janssen, 92190 Meudon, France 
              \email{rosine.lallement@obspm.fr}
  \and
  Univ. Grenoble Alpes, CNRS, IPAG, 38000 Grenoble, France
\and
  ACRI-ST, Sophia-Antipolis, France
\and 
ISTerre, Université Grenoble Alpes, France
}

   \date{Received ; accepted }

 
  \abstract
{Gaia stellar measurements are currently revolutionizing our knowledge of the evolutionary history of the Milky Way. 3D maps of the interstellar dust provide complementary information and are a tool for a wide range of uses. We aimed at building 3D maps of the dust in the Local arm and surrounding regions. To do so, Gaia DR2 photometric data were combined with 2MASS measurements to derive extinction towards stars that possess accurate photometry and relative uncertainties on DR2 parallaxes smaller than 20\%. We applied to the individual extinctions a new hierarchical inversion algorithm adapted to large datasets and to a inhomogeneous target distribution. \textbf{Each step associates regularized Bayesian inversions along all radial directions and a subsequent inversion in 3D of all their results}. Each inverted distribution serves as a prior for the subsequent step and the spatial resolution is progressively increased. 
We present the resulting 3D distribution of the dust in a 6 $x$ 6 $x$ 0.8 kpc$^{3}$ volume around the Sun. Its main features are found to be elongated along different directions that vary from below to above the mid-plane: the outer part of Carina-Sagittarius, mainly located above the mid-plane, the Local arm/Cygnus Rift around and above the mid-plane and the fragmented Perseus arm are oriented close to the direction of circular motion. The more than 2 kpc long spur (nicknamed the \emph{split}) that extends between the Local Arm and Carina-Sagittarius, the compact near side of Carina-Sagittarius and the Cygnus Rift below the Plane are oriented along l$\sim$40 to 55 $\fdeg$. Dust density images in vertical planes reveal in some regions a wavy pattern and show that the solar neighborhood within $\sim$500 pc remains atypical by its extent above and below the Plane. We show several comparisons with the locations of molecular clouds, HII regions, O stars and masers. The link between the dust concentration and these tracers is markedly different from one region to the other.}

   \keywords{Dust: extinction; ISM: lines and bands;  ISM: structure ; ISM: solar neighborhood ; ISM: Galaxy}

   \maketitle
%

\section{Introduction}

Evolutionary models of the Milky Way require measurements of the spatial distribution of massive numbers of stars as well as their dynamical, physical and chemical properties, all quantities currently provided by the ESA satellite Gaia \citep{Gaia16b,GaiaDR2} and ground-based surveys. In this context, three-dimensional (3D) maps of the Galactic interstellar (IS) dust are an additional and mandatory tool since they allow de-reddening of stellar spectra and suppression of degeneracies between stellar temperature and dust reddening on the one hand, and, on the other hand, since they may shed important additional light on the star formation through the inter-play between star and IS matter. Fortunately, Gaia parallaxes and Gaia photometric data complemented by ground-based photometric data allow the construction of the required IS dust maps in parallel with the stellar studies.  3D mapping is based on the tomographic inversion of distance-limited IS absorption data for large numbers of targets distributed in space at known locations: Gaia parallaxes evidently provide the target locations and Gaia plus ground-based photometric measurements provide estimates of the reddening along each sightline. Independently of these evolutionary aspects, 3D maps of the Galactic dust are a general tool for many various purposes such as studies of foreground, environment or background to specific objects, models of light or particle propagation, etc.  

One form of 3D mapping of dust is the construction of radial profiles of color excess or extinction, sightline by sightline. The first map produced in such a way was based on Hipparcos \citep{Arenou92}. Later on, maps of radial profiles and their derivatives, i.e. reddening or extinction per unit distance, were built by \cite{Marshall06}, \cite{Chen13} and \cite{Schultheis14} based on  the Besan{\c c}on model of stellar population synthesis \citep{Robin12} and 2MASS, GLIMPSE and VVV photometry, respectively. \cite{Majewski11} derived individual stellar reddening and radial profiles by combining SPITZER/GLIMPSE and  2MASS. \cite{Berry12} used SDSS and 2MASS to constrain stellar SEDs and reddening profiles using reddening-free empirical spectra. Following a hierarchical Bayesian method devised by \cite{Sale12},  \cite{Sale14} used photometric data from the IPHAS Survey to simultaneously reconstruct stellar properties and extinction radial profiles at a 10$\arcmin$ resolution for -5$\leq$b$\leq$5$\fdeg$ and 30$\leq$l$\leq$210$\fdeg$ up to 5~kpc.  \cite{Green15,Green18} used 2MASS and Pan-STARRS~1 and a refined Bayesian method to derive reddening radial profiles at very high angular resolution (on the order of 7$\arcmin$) for the sky regions accessible to Pan-STARRS (0$\leq$l$\leq$240$\fdeg$ at low latitudes). Both \cite{Sale14} and \cite{Green15,Green18} used derivatives of extinction profiles to produce 3D maps of dust density. \cite{Kos14} used the RAVE spectrometric and photometric data and presented reddening and also DIB strength radial profiles for the longitude interval l=180 to l=60$\fdeg$. Recently \cite{Chen18} used Gaia DR2 parallactic distances and combined Gaia DR2, 2MASS and WISE photometry to derive extinction profiles along the Plane at high angular resolution (6$\arcmin$) by means of a new machine-learning algorithm trained on LAMOST, SEGUE and APOGEE datasets.

Full 3D inversions differ from the previous technique by imposing spatial correlations between volume densities of IS matter in all directions, therefore linking adjacent sightlines and deriving the volume density of the differential extinction everywhere in 3D space. The differential extinction at point P in 3D space (also call extinction density) is the amount of extinction per unit distance suffered by stellar light traveling in the region enclosing P, here expressed in magnitude per parsec. In the case of the nearby ISM, three-dimensional inversions following a technique developed by \cite{Vergely01} have been applied to color excesses from ground-based photometry \citep{Vergely10, Lallement14} or  composite sources combining photometry and DIBs \citep{Capitanio17,lall_stilismfinal}.  \cite{SaleMagorrian14,SaleMagorrian15,SaleMagorrian17} developed a Gaussian field method adapted to realistic multi-scale IS matter distributions. \cite{Rezaei17, Rezaei18} developed a non-parametric 3D inversion method based on an isotropic Gaussian process to derive the 3D dust distribution and tested then applied their new technique on extinction-distance measurements derived from the APOKASC catalog  of  \cite{Rodrigues14} and APOGEE data respectively. 
Section \ref{Extinctions} describes how extinctions are estimated based on the Gaia and 2MASS photometric measurements and how various selection criteria and filters have been applied. Section \ref{inversions} details the hierarchical inversion technique, its limitations and the chosen parameters and presents the resulting multi-resolution map by means of several images of the dust density in various selected planes. Section \ref{Comparisons} displays comparisons  between the dust distribution within the Plane and  tracers of star formation and molecular clouds. Finally, section \ref{Discussion} discusses perspectives for 3D dust mapping. The appendix shows some comparisons between reddening profiles obtained by integration through our 3D dust distribution and those derived by \cite{Green18} based on Pan-STARRS and 2MASS photometry.

\begin{figure}[t]
\centering
\includegraphics[width=8cm]{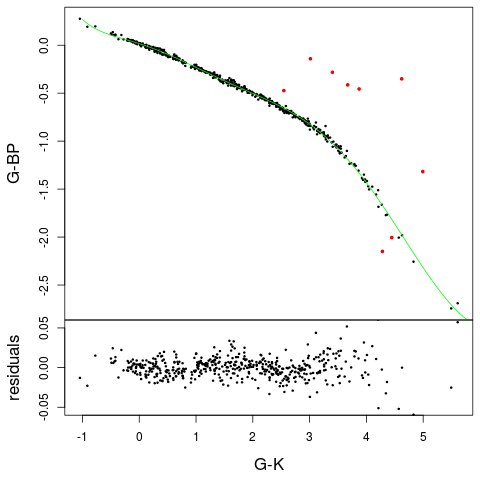}
\caption{Example of an intrinsic colour-colour relation fit (top) and its residuals (bottom) for $G-G_{BP}$ vs $G-K$. In red the outliers removed in the process.}
\label{fig:ccrel}
\end{figure}

\begin{figure}[t]
\centering
\includegraphics[width=7cm]{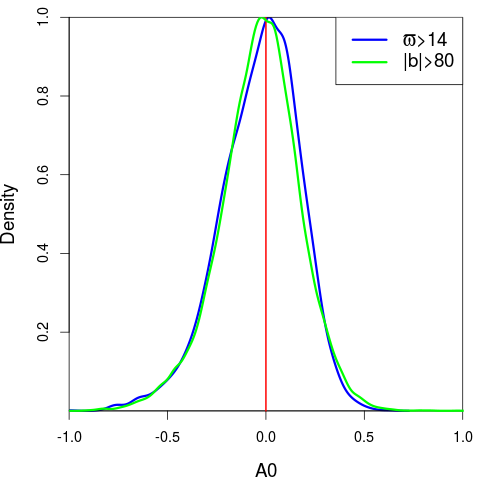}
\caption{A$_{0}$ extinction distribution (mag) in low extinction selections. In blue, the local bubble selection ($\varpi>14$ mas, corresponding to the first 70~pc); in green, a high latitude selection ($|b|>80\degr$). The median residuals are smaller than 0.03~mag \textbf{with a standard deviation of 0.2~mag}.}
\label{fig:lowexttest}
\end{figure}

\begin{figure}[t]
\centering
\includegraphics[width=7cm]{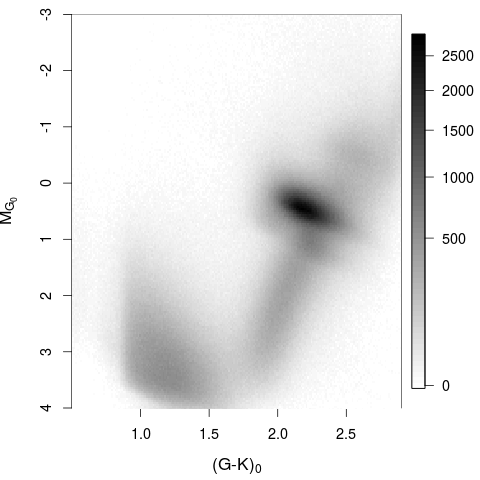}
\caption{Dereddened Hess diagram for a subset of stars with a parallax relative uncertainty smaller than 10\%. The grey scale corresponds to the square root of the stellar density. The Red Clump is the most prominent feature \textbf{which shape can be compared to Fig. 2 of \cite{RuizDern17}}.}
\label{fig:hrdtest}
\end{figure}

\begin{figure}[t]
\centering
\includegraphics[width=7cm]{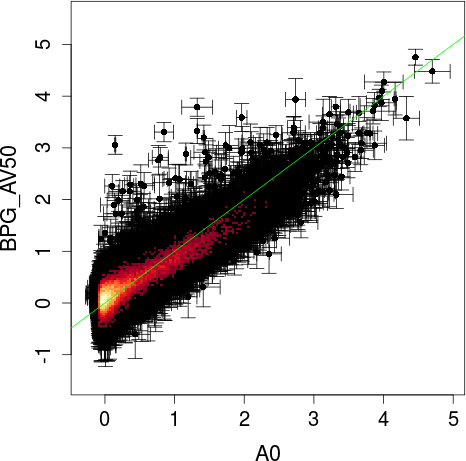}
\caption{Comparison between the extinction $A_0$ derived here and $A_V$ determined by \cite{Santiago16} using APOGEE data.}
\label{fig:apogeetest}
\end{figure}

\section{Deriving the extinction}\label{Extinctions}

To obtain the extinction, we use the intrinsic colour-colour relations and we model the extinction coefficients. This is done in the following way. To derive the extinction of individual stars, we combine the Gaia DR2 \citep{GaiaDR2} and the 2MASS \citep{2006AJ....131.1163S} photometry. We use the catalogue cross-match provided within the Gaia archive \citep{DR2-DPACP-41}. 
We applied the Gaia photometric and astrometric filters as provided in \cite{DPACP-31} (photometric uncertainties lower than 2\% in $G$, 5\% in $\BP$ and $\RP$, and filters on the photometric excess factor and an equivalent to the astrometric excess noise filter). We did not apply the filter on the number of visibility periods used as it was removing specific regions of the sky while the most critical outliers (negative parallax and too large parallaxes) are already removed by our other filters. We then selected stars with parallax relative uncertainties smaller than 20\%.  For 2MASS we selected stars with 2MASS photometric quality flag AAA and photometric uncertainties lower than 0.05 mag. 
Only stars on the top of the HRD ($M_G<5$) are selected to probe large distances and minimize the influence of binaries. 
The photometry has been handled as suggested by \cite{Weiler18}, with 2 different zero points, filters and photometric calibrations for the $\BP$ photometry for $G$ brighter and fainter than 11~mag and a correction for the $G$ magnitude drift \citep{Arenou18, Weiler18, CasagrandeVandenBerg18} of 3.5~mmag per magnitude. Stars brighter than $G<6$~mag have been removed to avoid saturation issues and fainter than $\BP<18$ to avoid background subtraction issues \citep{Evans18, Arenou18}. Following all the criteria, the total number of selected objects is $\sim$27,340,000.  

The photometric calibration has been performed on low extinction stars selected to have $E(B-V)<0.1$~mag using the previous 3D extinction map of \cite{Capitanio17}. 
A 7 degree monotone polynominal\footnote{R package MonoPoly} is adjusted to the colour-colour relations $G-X$ as a function of $G-K$, removing one by one strong (10 times the median absolute deviation (MAD) of the residuals) outliers, as illustrated in Fig.~\ref{fig:ccrel}. The bright stars calibration is done using stars with $G<10.5$ and is applied to all stars with $G<11$~mag.  The colour range of our calibration is $-1.0<\GKO<5.6$.

The extinction coefficients are derived using the \cite{FitzpatrickMassa07} extinction law and the Kurucz Spectral Energy Distributions \citep{CastelliKurucz03}, as described in \cite{Danielski18}. The 2MASS transmissions are taken from \cite{Cohen03} and the Gaia transmissions are those of \cite{Weiler18}. We are selecting here only the top of the HRD. We therefore adapted the surface gravity of the Kurucz spectra to the temperature with $\logg = 4$ for $\Teff>5250$ and $\logg = -8.3 + 0.0023~\Teff$ for cooler stars \textbf{(simple fit adjusted on APOGEE \citep{Majewski17} data)}. We also increased the degree of the polynomial fit of the extinction coefficient model:

\begin{equation}
k_m = a_1 + a_2 C + a_3 C^2 + a_4 C^3 + a_5 A_0 + a_6 A_0^2 + a_7 A_0^3 + a_8 A_0 C  + a_9 A_0 C^2 + a_{10} C A_0^2 
\end{equation}

with $C=\GKO$. The fit was performed on a grid with a spacing of 250~K in $\Teff$ with $3500<\Teff<1000$~K and a step linearly increasing by 0.01~mag in $A_0$ with $0.01<A_0<20$~mag. This increase of the step size with $A_0$ is made to ensure the best model of the extinction coefficients at low extinction values which dominates our sample. The residuals are smaller than 0.6\% for $A_0<10$~mag in the $G$ band. 

We tested the influence of the Gaia passband used to derive the extinction coefficients on our results and see only small differences between the revised Gaia DR2 passbands provided by \citep{Evans18} and those of \cite{Weiler18}. 

We pre-selected intrinsically bright stars using the 2MASS $Ks$ magnitude, less affected by extinction:
\begin{equation}
 K_s + 5 + 5 \log_{10} \left( \frac{\varpi + \varsigma_\varpi}{1000} \right) < 4
\end{equation}

where $\varpi$ is the Gaia DR2 parallax. \textbf{This cut allows to retrieve all stars with $M_G<5$ up to 5~mag of $\AO$ extinction, which covers the range of most of our extinctions.}
For each star the extinction $A_0$ and the intrinsic colour $(G-K_s)_0$ and their associated uncertainties are determined through a Maximum Likelihood Estimation\footnote{R package bbmle}. 

\begin{equation}
\mathcal{L} = \prod_X P(G-X | A_0, (G-K)_0)
\end{equation}

for $X=G_{BP}, G_{RP}, J, H$. To avoid local minima, 3 different initial values are tested : $(G-K)_0=G-K$ (no extinction), $(G-K)_0=1.5$ and $(G-K)_0= (G-K)_\mathrm{max}$ (the maximum colour of our intrinsic colour relation, e.g. 5.6~mag). We use the intrinsic colour-colour relations to derive $(G-X)_0$ from $\GKO$. Then $P(G-X | A_0, (G-K)_0) = P(G-X | (G-X)_0 + k[A_0,(G-K)_0] A_0)$. We model this probability by a Gaussian, quadratically adding the photometric error in the $X$ band and the intrinsic scatter of the colour-colour relation. 
Negative values of $A_0$ are allowed, to ensure a Gaussian uncertainty model, needed for the inversion method. However no extrapolation of the extinction coefficients is done, the negative values of $A_0$ being replaced by 0 to derive the extinction coefficients. 
Typical resulting uncertainties are 0.3~mag in $A_0$ and 0.2~mag in $\GKO$. 

A chi-square test is performed to check the validity of the resulting parameters, removing stars with a p-value limit smaller than 0.05. 
We removed stars with uncertainties on the derived $A_0$ and $\GKO$ higher than 0.5 and 0.4~mag respectively, as well as \textbf{outliers} for which $A_0$ is not compatible with being positive at 3 sigma. 
To ensure that we lie within our calibration interval, we keep only stars with dereddened $M_{G_0}< 5$.
Inspection of the de-reddened HRD shows that the bluest and reddest stars are not correctly recovered. We therefore also removed stars with $\GKO<0.5$ and $\GKO>2.9$~mag. 


We tested our individual extinctions by checking the extinction distribution in low extinction regions (Fig.~\ref{fig:lowexttest}) and the shape of the de-reddened HRD (Fig.~\ref{fig:hrdtest}). We also compared our extinctions with previous work, Fig.~\ref{fig:apogeetest} showing the comparison with extinctions derived from APOGEE spectroscopic data by \cite{Santiago16}.



\begin{figure*}[t]
\centering
\includegraphics[width=0.32\textwidth]{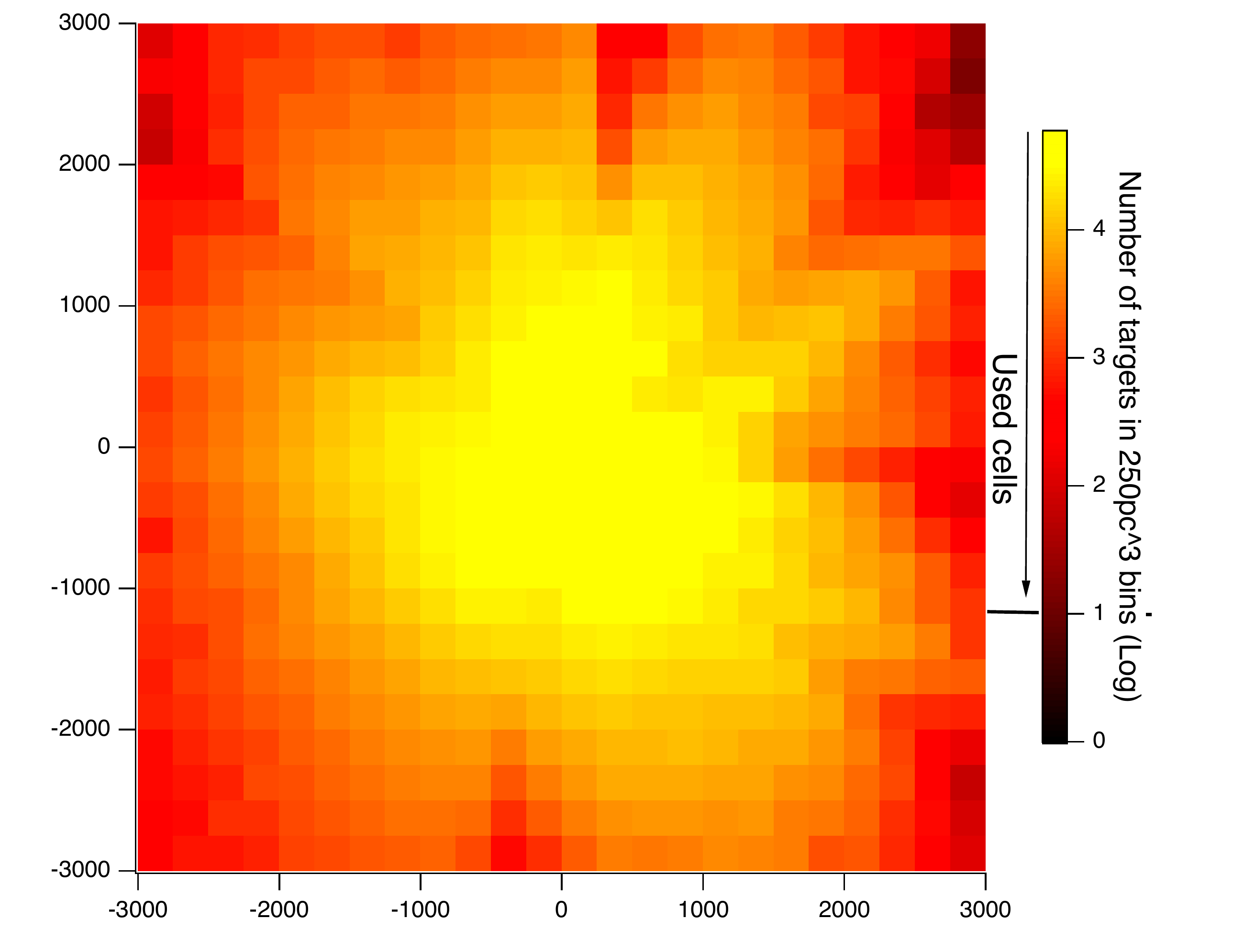}
\includegraphics[width=0.32\textwidth]{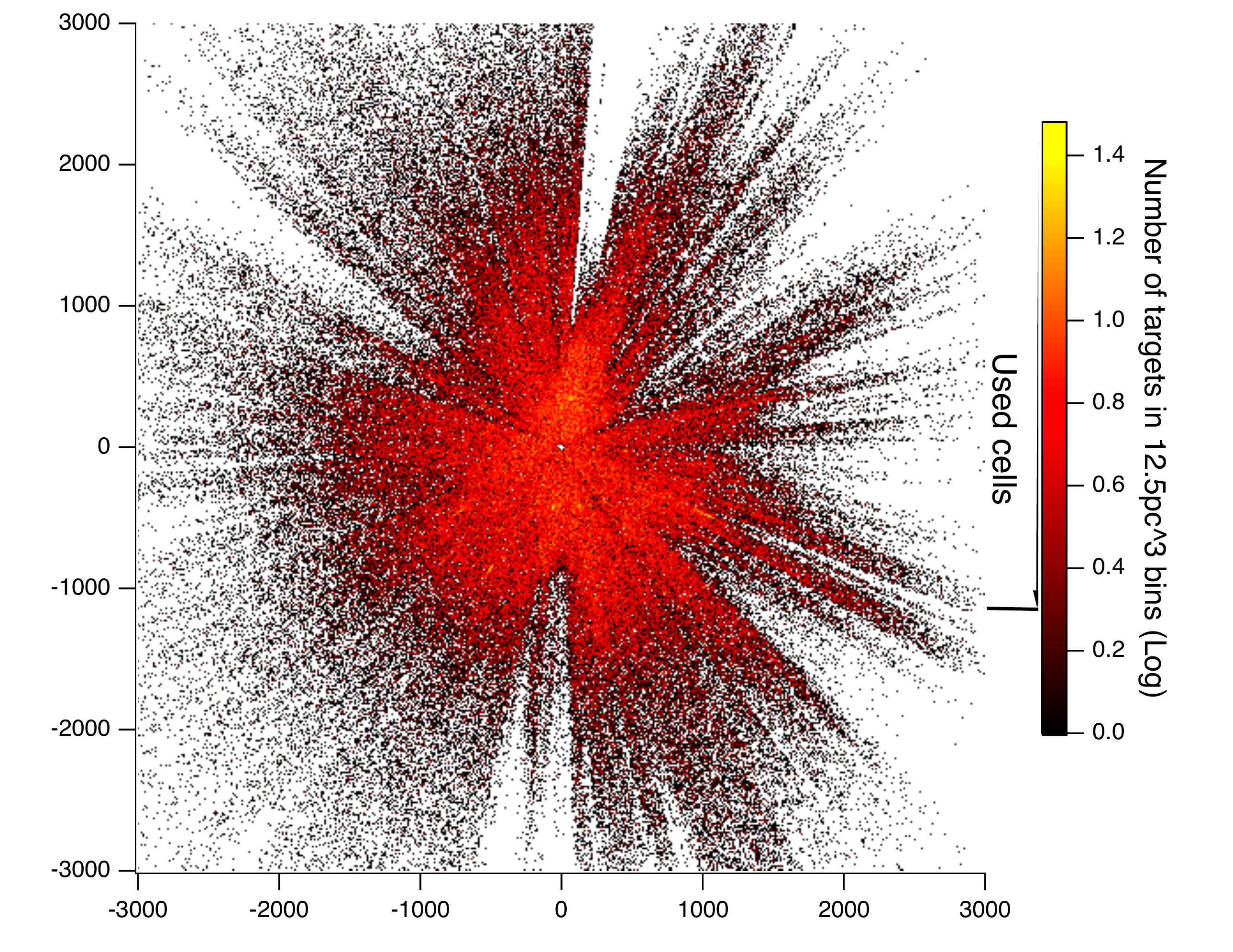}
\includegraphics[width=0.32\textwidth]{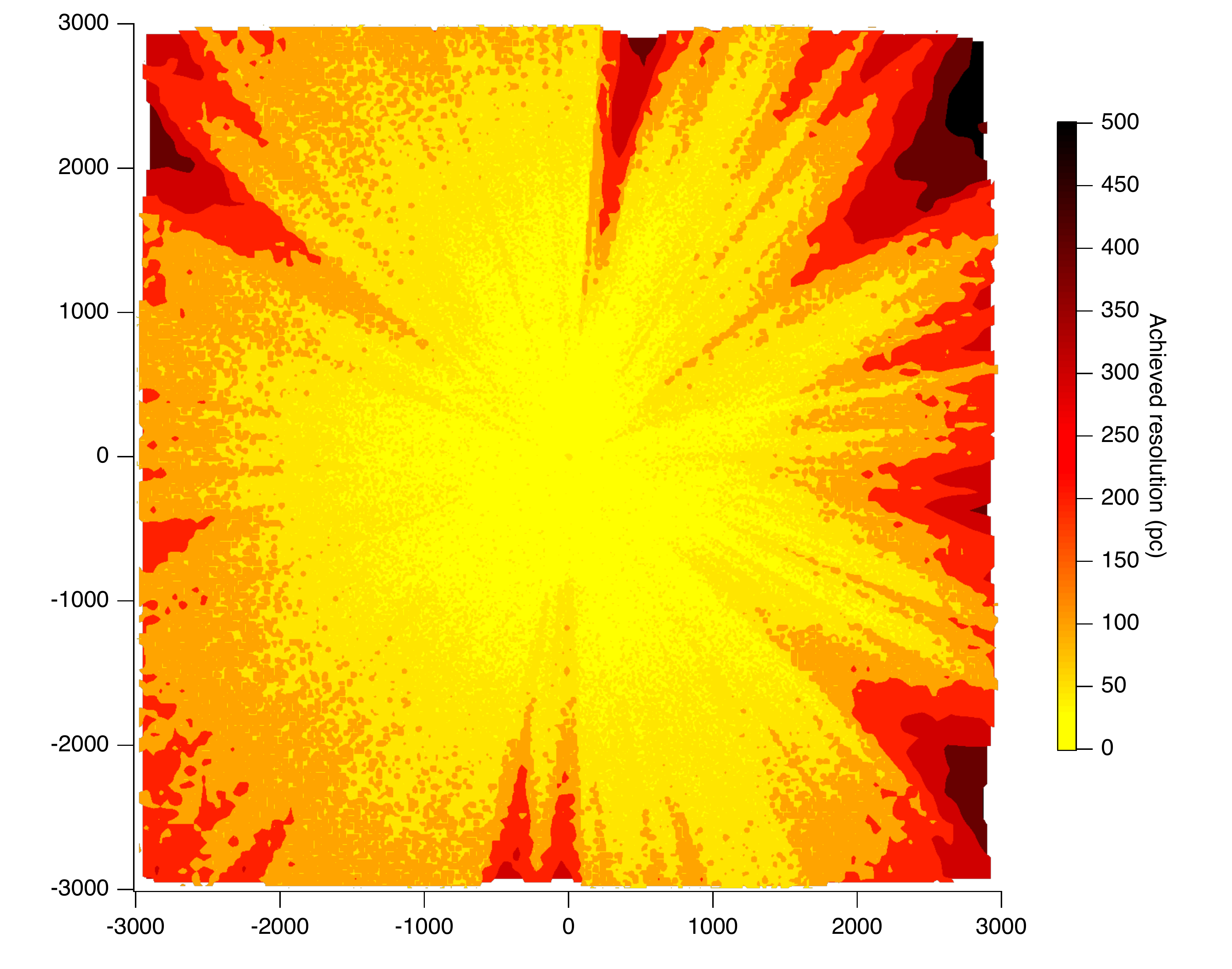}
\caption{Left: 250x250x250 pc bins \textbf{similar in volume to those} used in the first step of the inversion. Shown are the cells centered on the Plane. The Sun is at the center of the figure at (0,0). The Galactic center is to the right. The color code corresponds to the number of target stars per bin, \textbf{i.e. a quantity proportional to the target space density}. Only bins with more than 10 targets are used at this stage. Middle:  12.5x12.5x12.5 pc bins used in the last step of the inversion. Only bins with more than 2 targets are used. Right: Resolution achieved in the Plane based on the hierarchical inversion. In regions of high target \textbf{space} density, essentially close to the Sun, the resolution is 25 pc and is achieved after the last step. In regions of scarce targets, mainly at large distances, the resolution is coarse, may reach 500pc, and is achieved at the first steps. Such regions are not improved during the following steps.}
\label{fig:binsizes}
\end{figure*}

\begin{figure*}[t]
\centering
\includegraphics[width=0.45\textwidth]{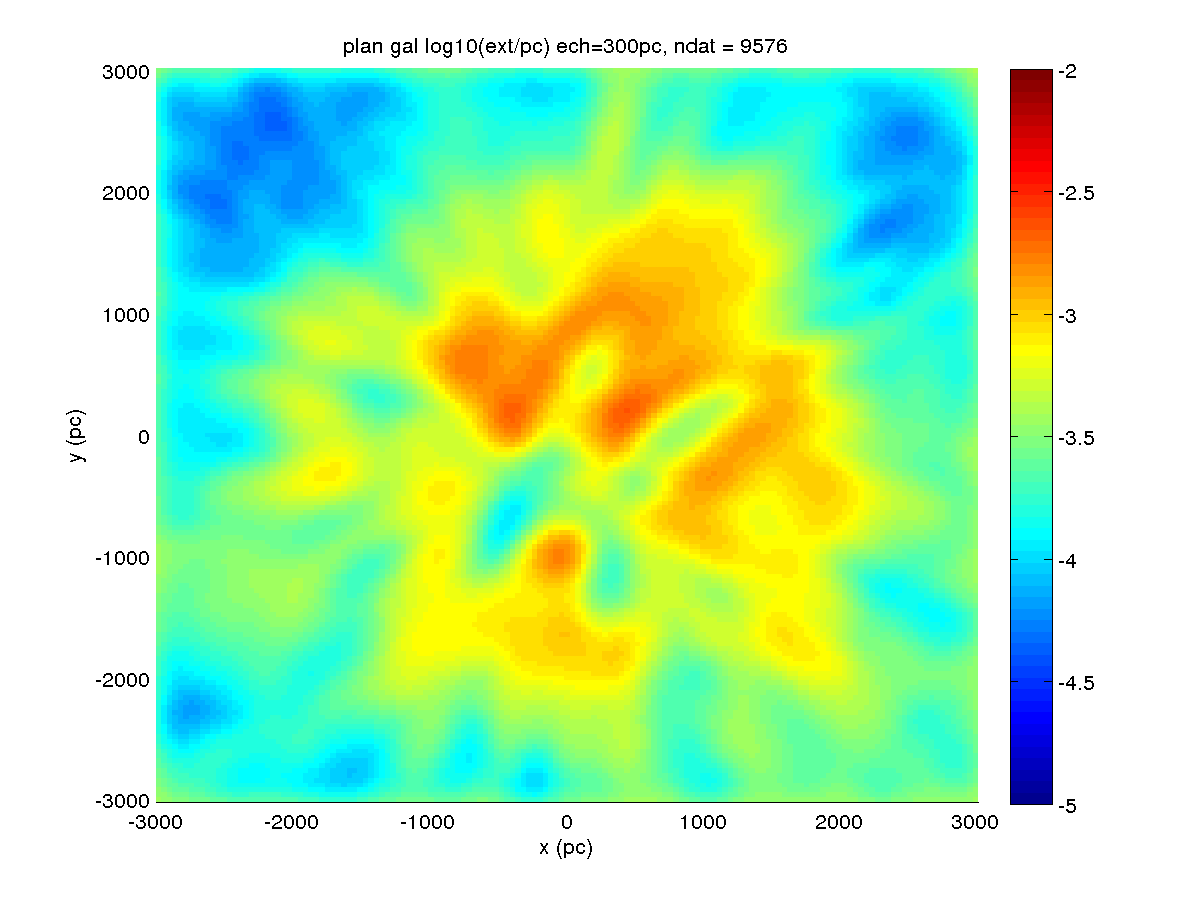}
\includegraphics[width=0.45\textwidth]{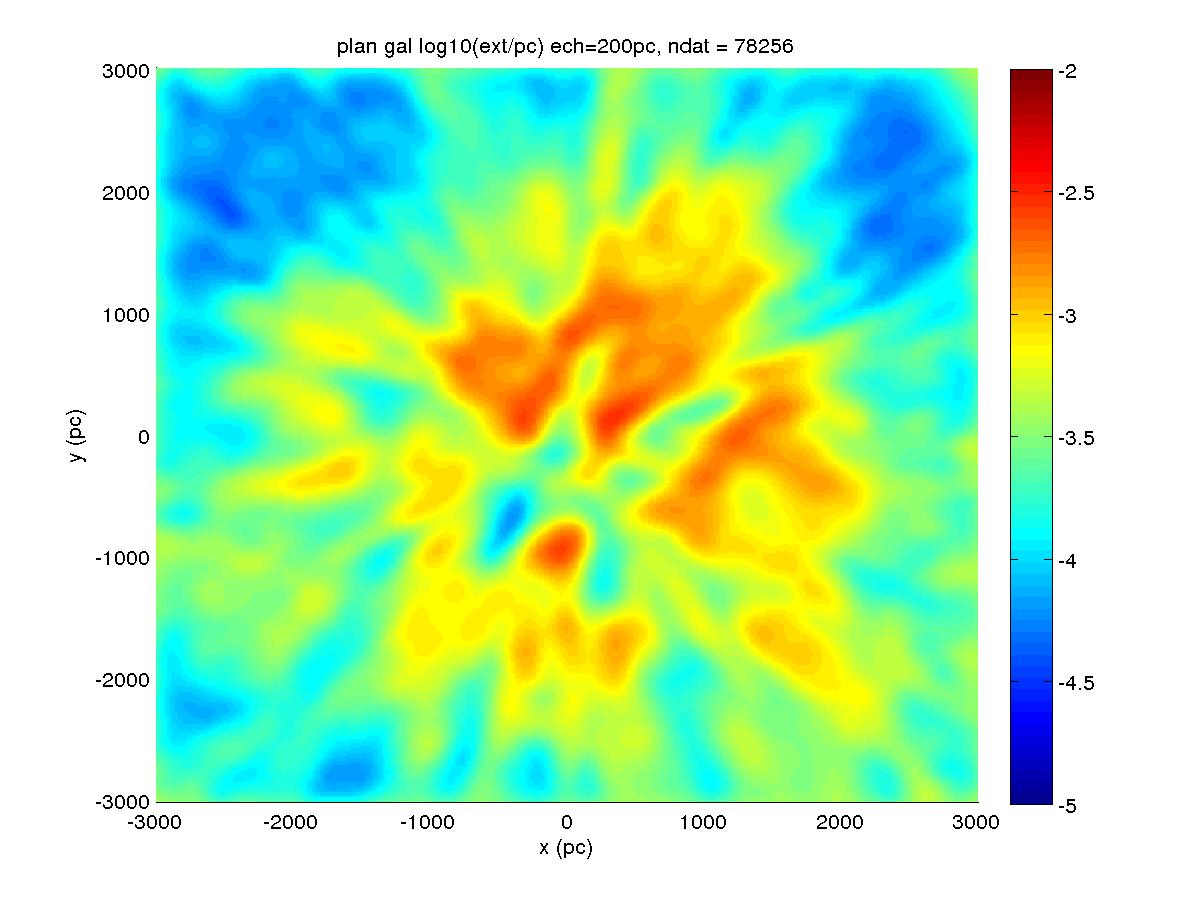}
\includegraphics[width=0.45\textwidth]{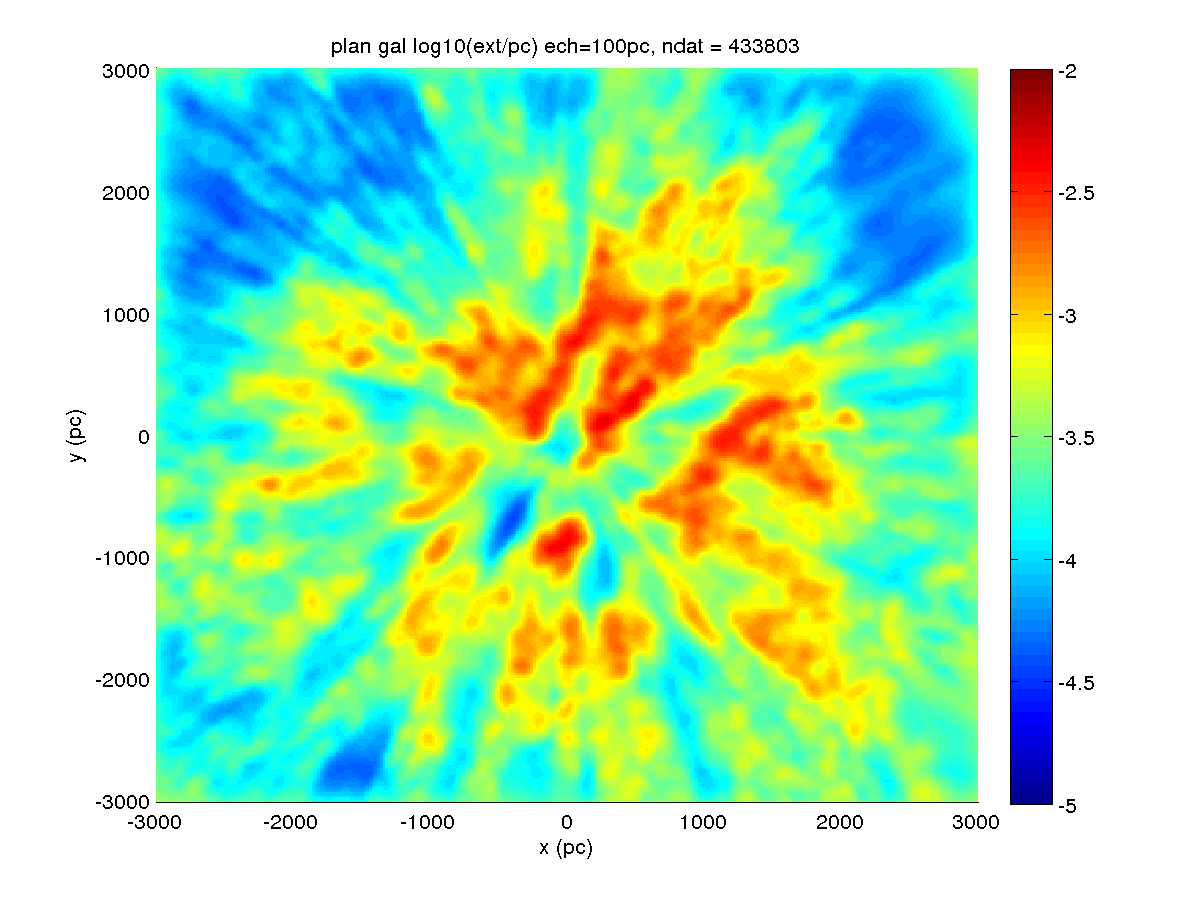}
\includegraphics[width=0.45\textwidth]{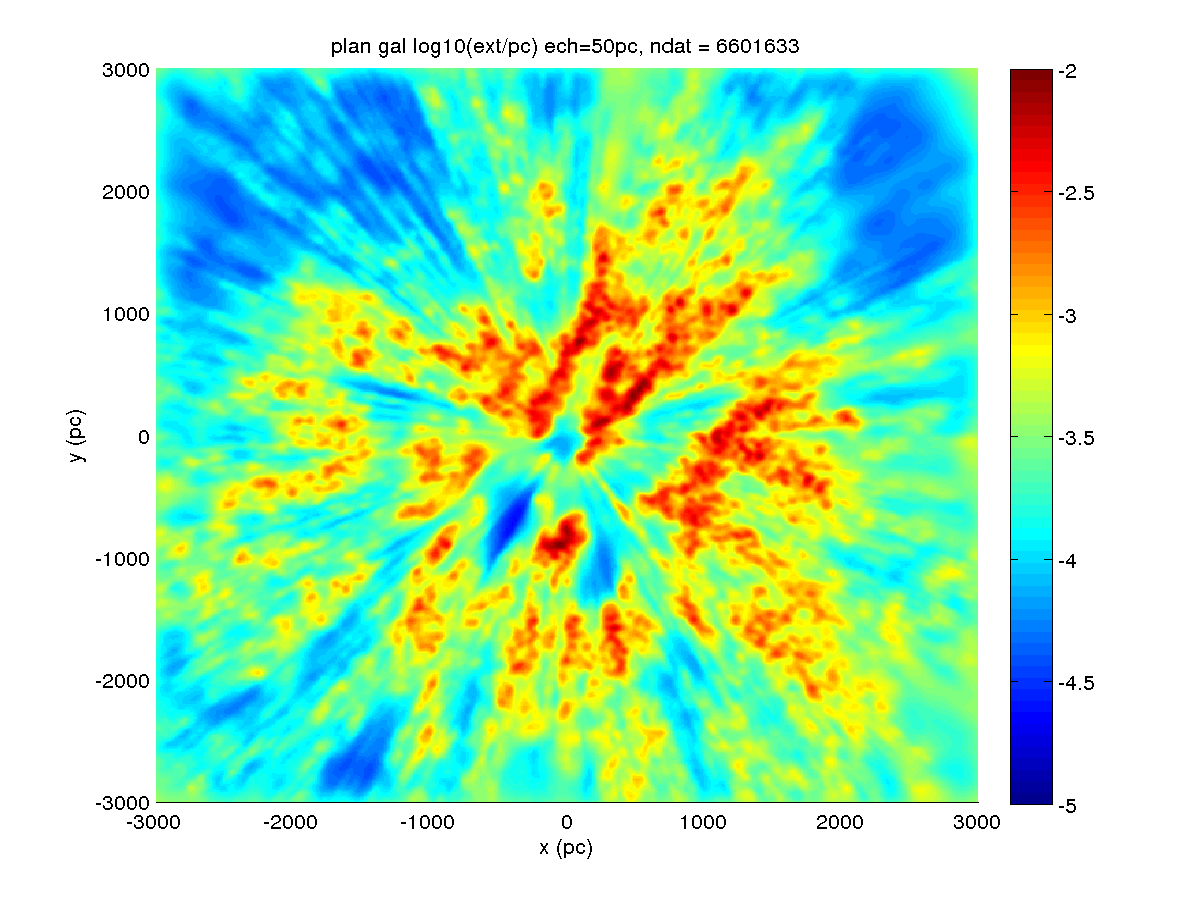}
\caption{Inverted 3D distribution of differential extinction, along the Galactic plane and as it is obtained after each of the first four iterative steps. The Sun is at the center of the figure at (0,0). The Galactic center is to the right. The achieved resolution is 300, 200, 100, and 50 pc respectively.}
\label{fig:hierarchic}
\end{figure*}

\section{Inversion of extinctions}\label{inversions}

\subsection{Description of the inversion technique}
The inversion method used to estimate the local differential extinction (in mag.pc$^{-1}$ units) is derived from the one presented in section 3 of \cite{lall_stilismfinal}. It was initially developed for the construction of a 3D distribution of local dust, to be used as a complement to large scale models in a new version of the GUMS model \citep{Robin15, Robin12}. 

The basic principles of the method are as follows. The treatment is hierarchical from the point of view of the spatial resolution (or structure scale size), with increased resolution at each iteration.  For each considered resolution, it is carried out in 2 steps: a first step consists in estimating the extinction along radial directions by means of a robust Bayesian inversion of individual extinctions, then a second step is a full 3D inversion of the first step results to produce differential extinction in 3D space at the resolution corresponding to the current hierarchical level. The covariance functions for the two radial and 3D steps are adapted to the current resolution. At variance with previous works based on \textbf{a covariance function that was a combination of two different functions} and aiming at better representing both dense and more diffuse clouds, here a unique Gaussian covariance function is used in the inversions. \textbf{The full width at half maximum of the Gaussian kernel is equal to chosen scale and is subsequently decreasing along the iterative steps}. The hierarchical approach makes it possible to process a very large volume of data with strongly varying spatial  density of targets \textbf{(i.e. number of stars per unit volume)} entering the inversion, while ensuring coherence of the information at a given scale.

The iterations proceed by progressively refining the solution (or equivalently the resolution), i.e., this approach makes it possible to manage the larger scales first and then go to the finest scales. Since the coverage of the data is not homogeneous, the spatial resolution obtained in the final solution varies from one place to another. In order to estimate the extinction on a given spatial scale, we average extinction data in boxes organized along radial beams, with a box size d$_{b}$ corresponding to half of the considered scale size, i.e., if we consider the 200 pc scale, we compute the average extinction in boxes of 100 pc size using all the stars contained in each box. \textbf{The boxes and the decomposition in radial beams are defined in the following way: the beam angle is the angle at which a segment the size of the box size  d$_{b}$ is seen from the Sun at mid-distance between the Sun and the extremity of the computational volume in the given direction. A box at distance d along the beam  is a truncated cone whose summit angle is this previously defined angle and whose thickness along the radial direction is d$_{b}$. The decomposition of the $4\pi$ sphere in beams and the distribution of box centers along the radial radiation are such that boxes overlap by 50\% in both radial and azimuthal directions} to ensure a consistency between average extinctions of two adjacent boxes. If the number of stars in a box is less than a chosen scale-dependent limit, then the information from the box is not taken into account in the inversion. Here our series of scales is: 500pc, 400pc, 300pc, 200pc, 100pc, 50pc, 25pc. The minimum number of stars required in a box depends on the scale and is: 10, 10, 10, 10, 5, 3, 2 for the 500pc, 400pc, 300pc, 200pc, 100pc, 50pc, 25pc scales respectively. Fig \ref{fig:binsizes} shows the decomposition in boxes  \textbf{with volumes similar to mid-distance boxes} used for the first and the last iterative steps, along the Galactic plane. The number of targets in each bin is color-coded, allowing to figure out which cells are used  (lower limit of 10 (resp. 2) targets for the first (resp. the last) iteration). The right panel shows  the  final resolution achieved in the Plane that results from the cell sizes and the minimum numbers of targets  at the various steps. As expected, the resolution is mainly governed by the radial distance. Note, however, several regions where it is particularly poor, e.g. for Galactic longitudes 25 or 85 $\fdeg$. We will come back to this point. 

More precisely, for each scale $n$, the inversion algorithm is as follows:
\begin{itemize}
\item
1) We start by calculating the limits of the boxes of scale $n$ along beams covering the whole sky. The angular aperture of the beams as the radial sampling along each beam depends on the scale.
\item
2) Then, we assign to all target stars their boxes and calculate the average extinction in each box with a weighting based on the individual errors. Moreover, the standard deviation of the extinction in each box is calculated from the variability of the extinction inside the box, i.e., this variability is considered to be an extinction error on the computed average.
\item
3) For each of the radial beams, the extinction density is inverted using the Bayesian method presented in \cite{Vergely10}. The beams are processed independently of each other. The auto-correlation function is a Gaussian kernel whose correlation length corresponds to the scale considered. An important point is the choice of the prior: at each stage the prior distribution is the interpolated 3D extinction density obtained at the previous stage of scale $n-1$.
\item
4) Extinction densities obtained along all different beams are assembled. An error is estimated based on the standard deviation (std) calculated among the nearest neighbor extinction densities. It is the std divided by the square root of N, where N is the number of neighbors.
\item
5) The extinction densities from all different beams are then inverted in 3D to update the 3D distribution obtained at the $n-1$ scale and  produce a 3D distribution at the scale $n$. The 3D density distribution obtained at the $n-1$ scale is used as a prior distribution for the inversion. To do so, the update of the cube is done by an optimal nonlinear interpolation method which imposes the positivity of the solution and which integrates a 3-sigma filtering (a data point which, after the first iteration, deviates from 3 sigmas of the model is discarded). The auto-correlation function used in the 3D inversion is the same as that of step 3. 
\end{itemize}
At the first step $n=1$ (or first spatial scale of 400 pc), the 3D extinction distribution entered as a prior is a homogeneous model with exponential decay on both sides of the Galactic plane, with a characteristic height of 200 pc. Extinction data are those described in the previous section for which distance from the solar plane Z is below or equal to 400 pc and distance along the  radial axis X and rotation direction Y is smaller than 3 kpc ($\sim$16,130,000 objects). Distances are derived from Gaia DR2 parallaxes, after subtraction of the zero point -0.03 mas \citep{Arenou18, Lindegren18}.

\subsection{Limitations and biases of the inversion}

The present hierarchical inversion has suppressed or at least strongly attenuated what was the main limitation of our full 3D mapping technique (used recently by \cite{Capitanio17, lall_stilismfinal}), namely the \textbf{use of a unique correlation length. As a matter of fact,  this single length could not be adequate simultaneously for low and high target space density, i.e., in general, for small and large distances. If the correlation length was chosen too wide, then some information at small distances was not used. On the contrary, if it was chosen too narrow, structures in distant regions were unrealistic.} Still, the best 3D spatial resolution associated with the last step is limited by our target space density and, e.g., does not allow yet reaching the scale of the filaments revealed by 2D maps or the very high angular resolution of the \cite{Green18} reddening profiles, as shown by the comparisons with Planck in Fig \ref{fig:galplaneintegr} (see below).
 However, in principle, iterative steps can be further continued to achieve better resolution without the introduction of artefacts far from the Sun. Searches for the lowest achievable size are in progress, and will benefit from future Gaia data releases. \textbf{This optimal size depends primarily on the spatial density of targets, but also on distance and extinction uncertainties.}

A bias of this inversion is the use of distances that are directly obtained from Gaia parallaxes, at all iterative steps. I.e., distances are simply calculated as the inverse of the DR2 parallax, after correction for the zero point of -0.03 mas. It is well-known that using this simple relationship underestimates actual distances. This has been studied in several ways and recently re-investigated by \cite{Luri18} in the context of Gaia DR2. In our previous maps, most of the inverted cloud complexes were at small distances (say, $\leq$ 1 kpc), and, more importantly, for the large majority of targets beyond several hundreds pc distances were photometric. Therefore, the problem was neglected. With the present dataset, we aim at mapping structures up to 3 kpc and all distances are parallactic. Therefore, biases may be significant and it is important to check potential consequences of our simplification. 

A first estimate can be made that is only based on the relative error on the parallax: following \cite{Arenou99},  if errors on the observed parallaxes are Gaussian,  then for small relative errors and in the absence of negative or null parallaxes the expected bias on the distance, i.e. the difference between the true distance and the expectation of distance using the inverse of the parallax $1/\varpi$  \textbf{, i.e. $E(d | \varpi = 1/\varpi$)} can be approximated as:
\begin{equation}
d - E(d | \varpi) \approx \frac{1}{\varpi} \cdot (\frac{\sigma(\varpi)}{\varpi})^{2}
\end {equation}

Having restricted the input source catalog to targets with relative errors on parallaxes lower than 20\%, or $\frac{\sigma(\varpi)}{\varpi} \leq 0.2$, we can assume that we are close to the small error regime and that this formula, that applies to the true parallax, can also provide an order of magnitude in the case of a measured parallax. Therefore, at the outer boundaries of our map (3 kpc along the X, Y axes in the Plane) and for the faintest targets, the bias in distance should be limited to a few hundreds of pc. 

\begin{figure}[t]
\centering
\includegraphics[width=14cm,height=12cm]{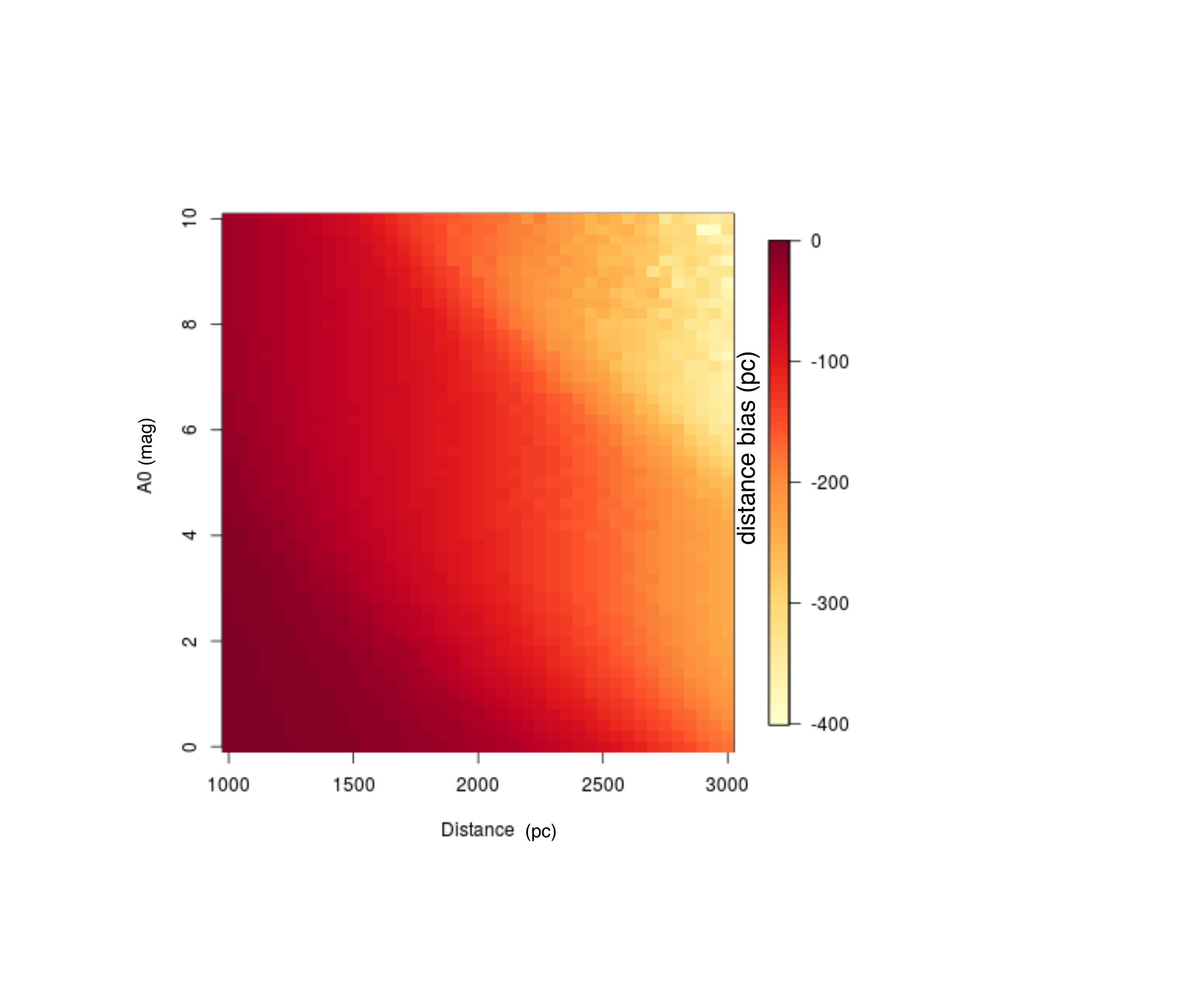}
\caption{Biases on the parallax distance, in pc, from a Monte-Carlo simulation adapted to our dataset}
\label{fig:bias}
\end{figure}

To estimate the distance bias in a way that takes into account not only the parallax error but also our target selections, we assumed that our method is equivalent to a weighted mean on the inverse of the parallax and simulated with a Monte-Carlo the Gaia errors on the Gaia HRD with $M_{G}>5$ shifted to different distances and extinctions. We applied the $\varpi/\sigma_\varpi>5$ criterion and simplified the magnitude cut to $G<20$. The resulting bias as a function of distance and extinction is presented in Fig.~\ref{fig:bias}. \textbf{The figure shows that, at fixed reddening, the effect of using $1/\varpi$ and selecting the sources as we made is to stretch the distance scale. Quantitatively,}  it is clear from the figure that, since most of our extinctions are lower than $\AO$=4, biases should be limited to 200~pc at 3~kpc, and less than 100~pc within 2~kpc. 

These estimates imply that for the large majority of our input data and at all distances, biases should be smaller than  the  resolution presently achieved (see Fig \ref{fig:binsizes}), e.g. biases of $\sim$ 150-200~pc at the outer boundaries of the map are smaller than the 300 to 500~pc (poor) resolution achieved at those distances.  
For this reason we neglected this effect for the present mapping. An \emph{a posteriori} correction of the maps is under study but is beyond the scope of this article.

As mentioned in previous work, the main advantage of our methods is the use of spatial correlations along all directions in 3D space to bring a global picture of the matter distribution. 
The price to pay is the lower limit on the sizes of the inverted structures. This shortcoming is increasingly important at increasing distances, and it is mandatory to understand its impact.  A first way to estimate the quality of the 3D distribution is illustrated in Fig. \ref{fig:galplaneintegr}. The top map is the latest 2D dust map from Planck, based on the optical thickness of the dust at 353 GHz \citep{Planck2016}. The level of details is extremely high, as expected. The middle 2D map represents the locations of all our input targets that are distant by more than 1~kpc. The color scale represents the reddening associated with each star.  It can be seen that the level of details reached by the Gaia-2MASS extinction dataset is also quite remarkable. The third 2D map shown at bottom is obtained through integration of the 3D distribution of differential extinction (taken from the last stage of the hierarchical inversion) along radial directions spaced every 0.25\fdeg\ in  longitude and latitude, from the Sun to the limits of our computational volume. It allows to figure out clearly which structures have been captured by the whole inversion process, how they have been smoothed, and which ones have been smoothed out. It can be seen that all major structures appearing in the first two maps are retrieved, however its level of details is inferior, as expected, and a few very small filamentary structures are no longer visible in this 2D representation. However, they have some impact on the 3D distribution that can be seen e.g. in images of the extinction density in selected planes.   

\begin{figure*}[t]
\centering
\includegraphics[width=10cm]{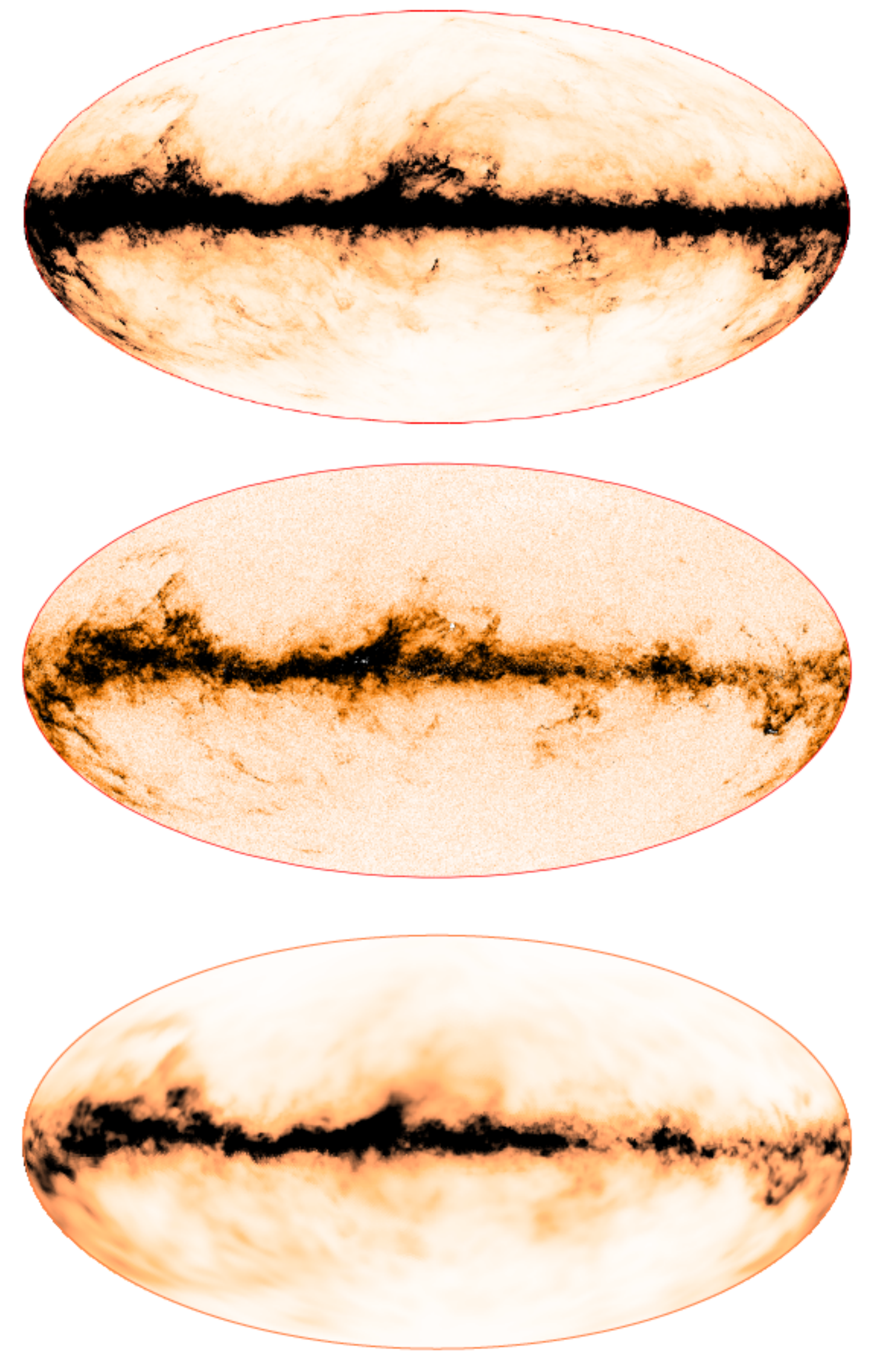}
\caption{Top: Planck map of the Galactic dust optical thickness $\tau$353 GHz. Middle: Estimated extinction towards all targets located beyond 1 kpc. Bottom: 3D differential extinction integrated from the Sun to the limits of the computational volume.}
\label{fig:galplaneintegr}
\end{figure*}

\begin{figure*}[t]
\centering
\includegraphics[width=15cm]{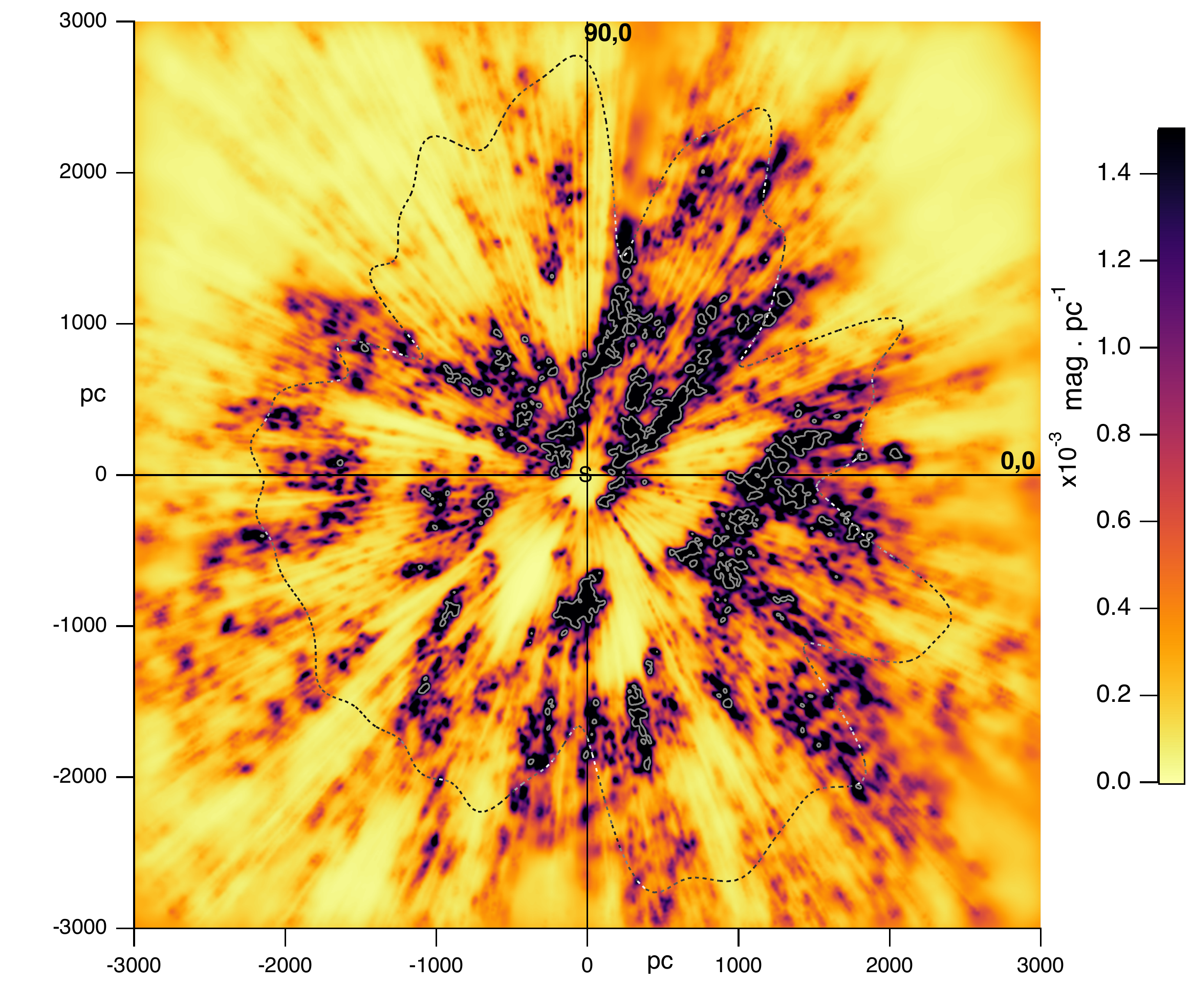}
\caption{Dust density along the Plane, based on the 3D distribution obtained after the last inversion step. The Sun is at the center of the figure at (0,0). The Galactic center is to the right. The color scale represents the differential extinction in units of mag per parsec. The dashed black line represents the distance beyond which the final resolution of 25 pc is not achievable due to target scarcity (see text). The dotted white contours correspond to a differential extinction of 0.003 mag.pc$^{-1}$ and delimit the dense areas.}
\label{fig:galplane}
\end{figure*}

\begin{figure*}[t]
\centering
\includegraphics[width=15cm]{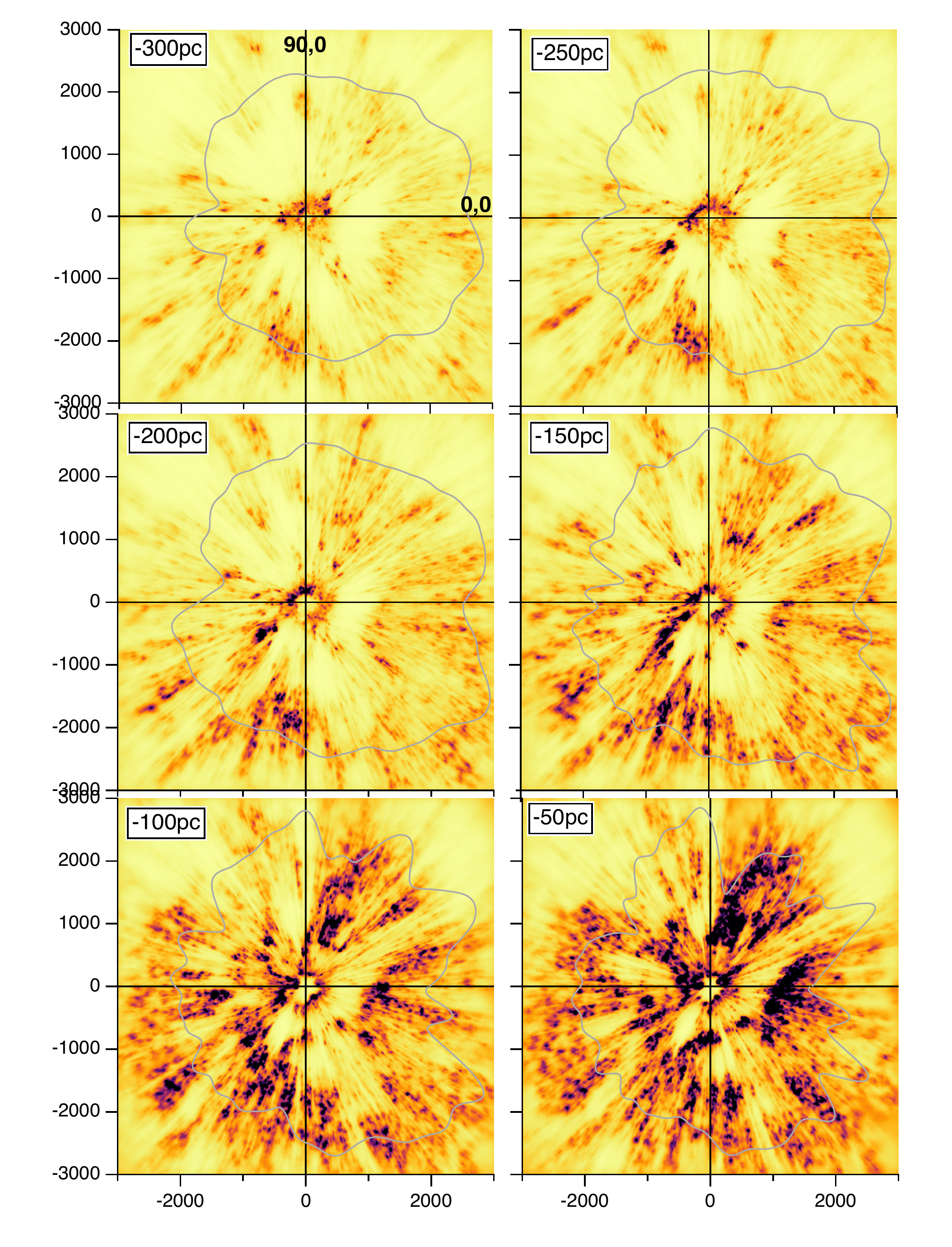}
\caption{Differential extinction in horizontal planes  300, 250, 200, 150, 100 and 50 pc below the Galactic plane. The color coding and the significance of the black line are identical to the ones in Fig. \ref{fig:galplane}}
\label{fig:south}
\end{figure*}

\begin{figure*}[t]
\centering
\includegraphics[width=15cm]{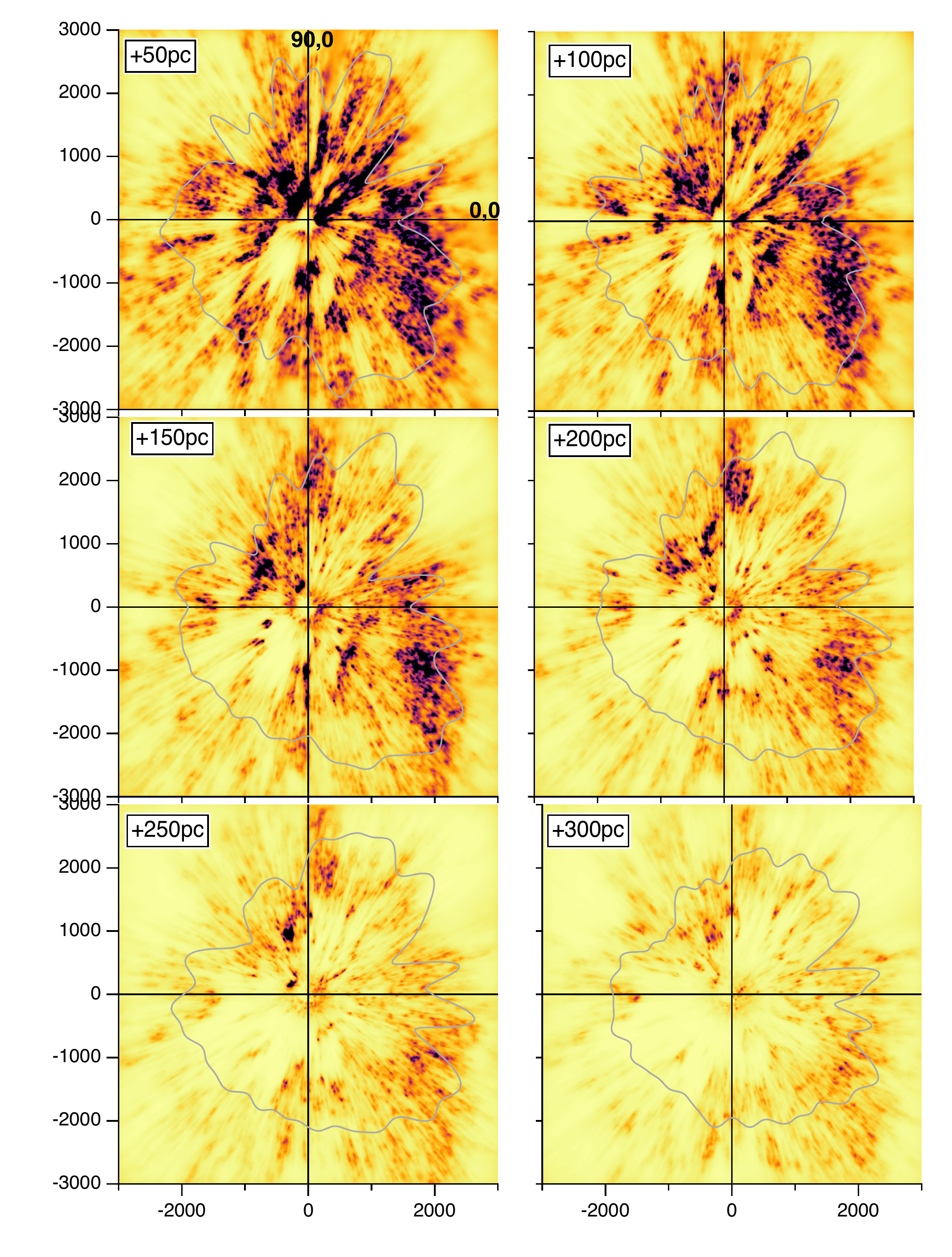}
\caption{Same as fig \ref{fig:south} for planes above the Galactic plane. Note the orientation of the Carina-Sagittarius region in the fourth quadrant, especially at Z=+100, +150pc, that differs by a large angle ($\sim45$ $\fdeg$) from the one of the compact, closer part seen in Fig. \ref{fig:galplane}.}
\label{fig:north}
\end{figure*}

\begin{figure*}[t]
\centering
\includegraphics[width=15cm]{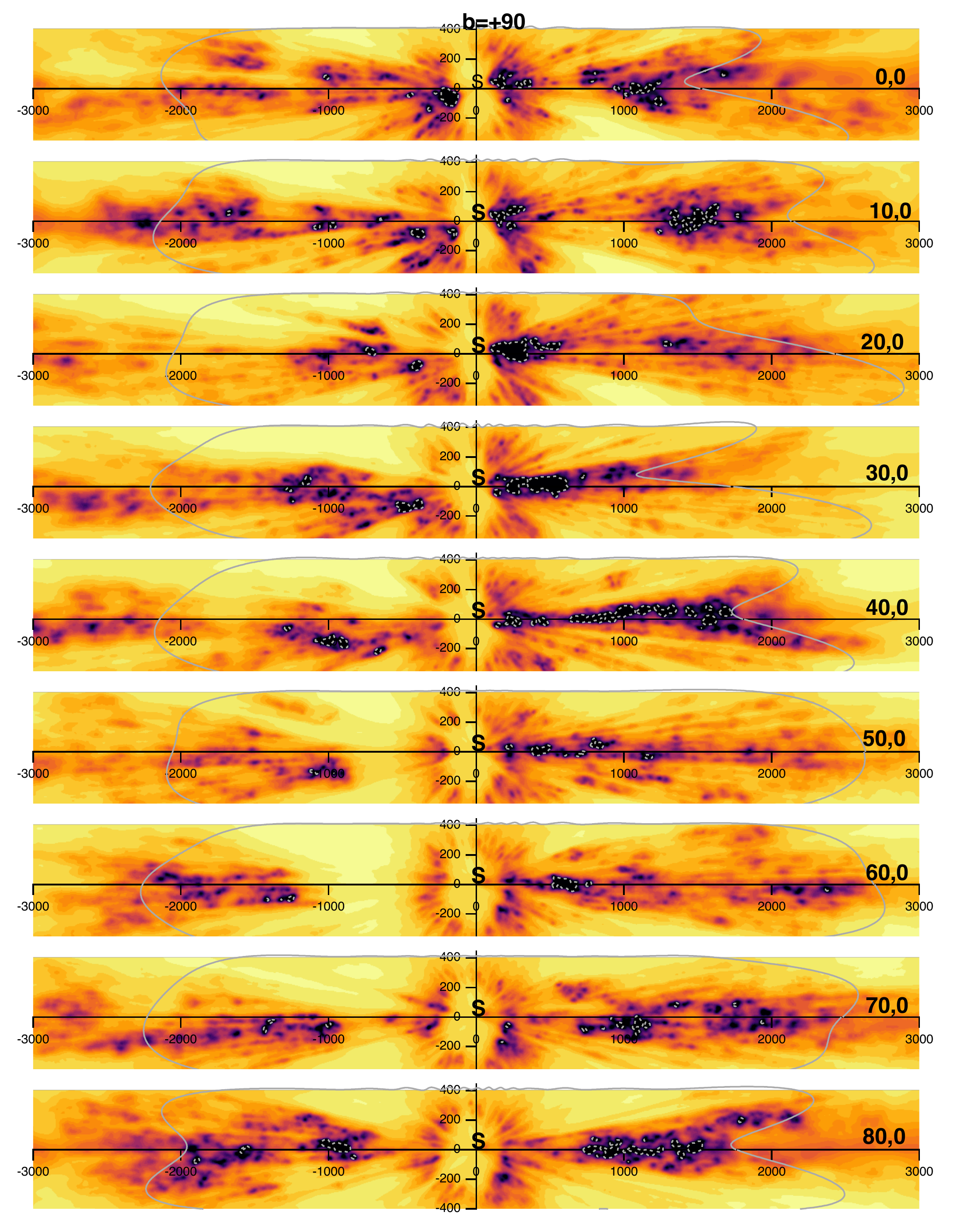}
\caption{Dust distribution in vertical planes containing the Sun and for longitudes spaced every 10$\fdeg$. The quantity represented is the square root of the differential extinction. The color scale is indicated on top of Fig. \ref{vertical2}. The dotted gray line represents the distance beyond which the final resolution of 25 pc is not achievable due to target scarcity (see text). The dashed light-grey contours correspond to a differential extinction of 0.003 mag.pc$^{-1}$. Note the wavy structure along the Plane for longitudes 200, 220, 250$\fdeg$.}
\label{fig:vertical1}
\end{figure*}

\begin{figure*}[t]
\centering
\includegraphics[width=15cm]{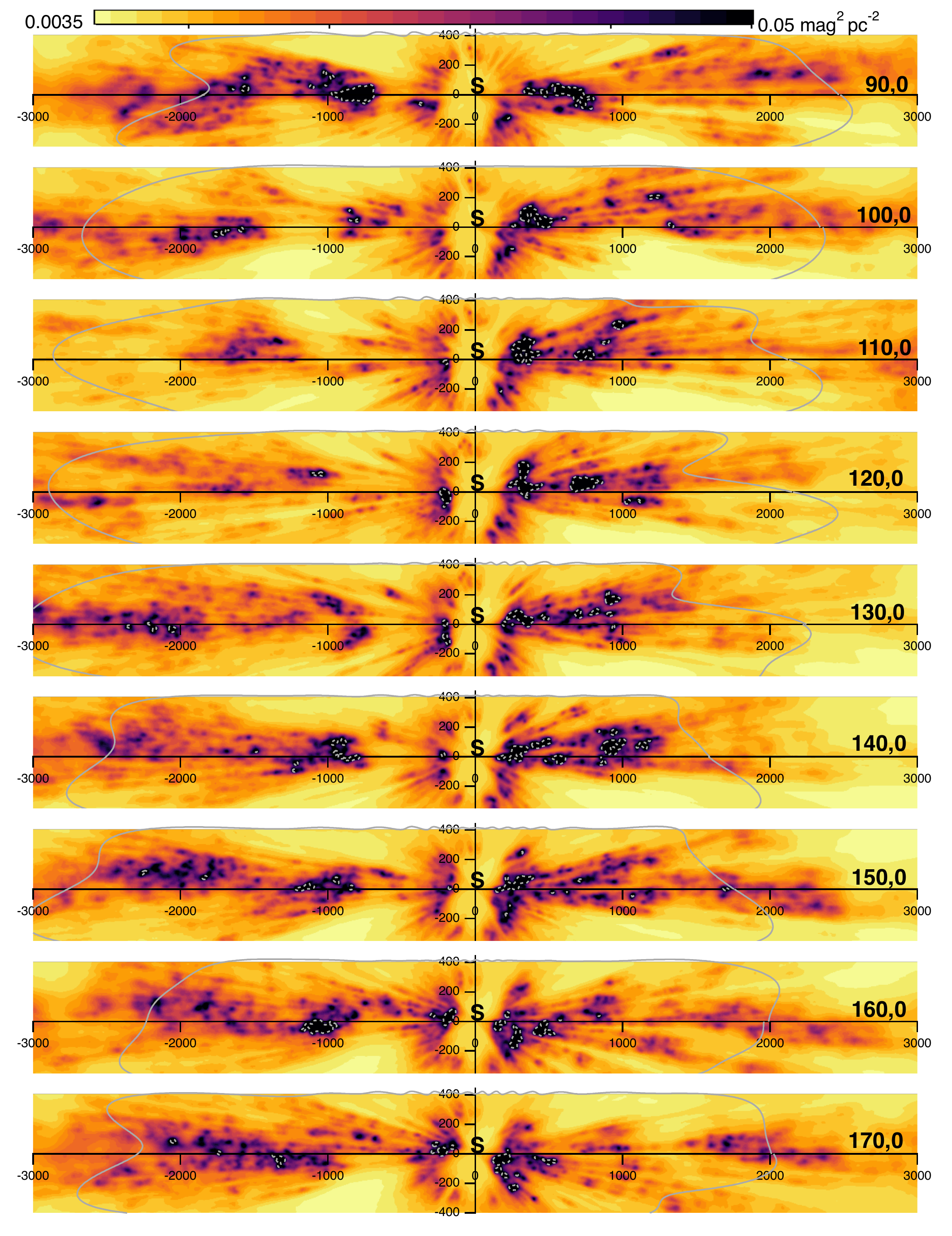}
\caption{Continuation of Fig \ref{fig:vertical1}. Note the wavy structure along the Plane, especially for longitudes 270$\fdeg$  and 330$\fdeg$.}
\label{fig:vertical2}
\end{figure*}

\section{Main features and comparisons with other maps and ISM tracers}\label{Comparisons}

Figures \ref{fig:galplane}, \ref{fig:south}, \ref{fig:north}, \ref{fig:vertical1} and \ref{fig:vertical2} display images of the extinction density in various horizontal and vertical planes. In each of these figures a white line separates regions where the target density is above or below the threshold of 2 per boxes chosen for the final resolution of 25 pc. I.e., regions beyond this limit suffer from target scarcity and a resulting poorer resolution and should be considered with more caution. The limit has been smoothed for a better visualization. The horizontal plane containing the Sun (i.e. parallel to the Galactic Plane and very close in distance, hereafter called solar plane or Plane) in Fig. \ref{fig:galplane} is particularly interesting since it covers a 6$x$6 kpc$x$kpc area and reveals the main nearby features. Figs. \ref{fig:south} and \ref{fig:north} display images of the dust density in horizontal planes at various distances from the solar plane, from Z=-300 pc to Z=+300 pc by steps of 50pc. Figs. \ref{fig:vertical1} and \ref{fig:vertical2} display images of the dust density in vertical planes containing the Sun and oriented along longitudinal axes varying by steps of 10 degrees.

It is interesting to compare the 3D distribution of dust with tracers of star-forming regions and molecular clouds (MCs). This is why, in addition to the density maps alone, we also display the locations of several tracers of star formation and of MCs, each time superimposed on the solar plane density map. In order to facilitate the visualization, we display the dust density in logarithmic scale and it is color coded in a different way, with reduced contrast.

\begin{itemize}
\item
Fig \ref{fig:masers_arms} displays the locations of nearby masers from \cite{Reid14}, \cite{Xu16} and \cite{Honma12} that possess parallax distances and are located within 50 pc from the Plane. In addition to the masers, the spiral arms that have been adjusted to the data by \cite{Reid14} and \cite{HouHan14} are displayed, i.e., here, the Perseus, Local and Carina-Sagittarius arms as well as the \emph{Local Spur} inferred by \cite{Xu16}.

\item 
Fig \ref{fig:planecomp_regions} shows the projections onto the Plane of the HII regions from the \cite{Russeil03} and \cite{HouHan14} catalogs.  The \cite{Russeil03} star-formation regions are deduced from H$\alpha$, H109$\alpha$, CO, radio continuum and absorption lines, and distances are based on parallaxes, or photometric or kinematic data, depending on the objects. Additional HII regions compiled by \cite{HouHan14} and that correspond to various transitions of H, He, CS, CO and more recent observations (see references in \cite{HouHan14}) are also represented. We restrict the figure to objects with distances estimated from stellar parallaxes and/or stellar photometric determinations and located within less than 50 pc from the solar plane. 

\item
Fig \ref{fig:xu_OBstars}  displays the O stars from the \cite{Reed2003} catalog  that have been very recently cross-matched for their distances with Gaia DR2 by \cite{xu_comparison_2018}  and that lie within 50 pc from the Plane.

\item
Fig \ref{fig:MCs} displays the \cite{miville-deschenes_physical_2016} molecular clouds derived from CO spectral-cube decomposition and kinematic models, completed by the fraction of the giant molecular clouds listed by \cite{HouHan14} that are located within 3 kpc. Again, only objects closer than 50 pc from the Plane are retained.

\end{itemize}

\subsection{Main structures}

 Hereafter we discuss the most prominent dust cloud structures and their degree of agreement with the various tracers, based on the images quoted above. Discussions about specific regions are beyond the scope of this article. 
 
\begin{figure}[t]
\centering
\includegraphics[width=9cm]{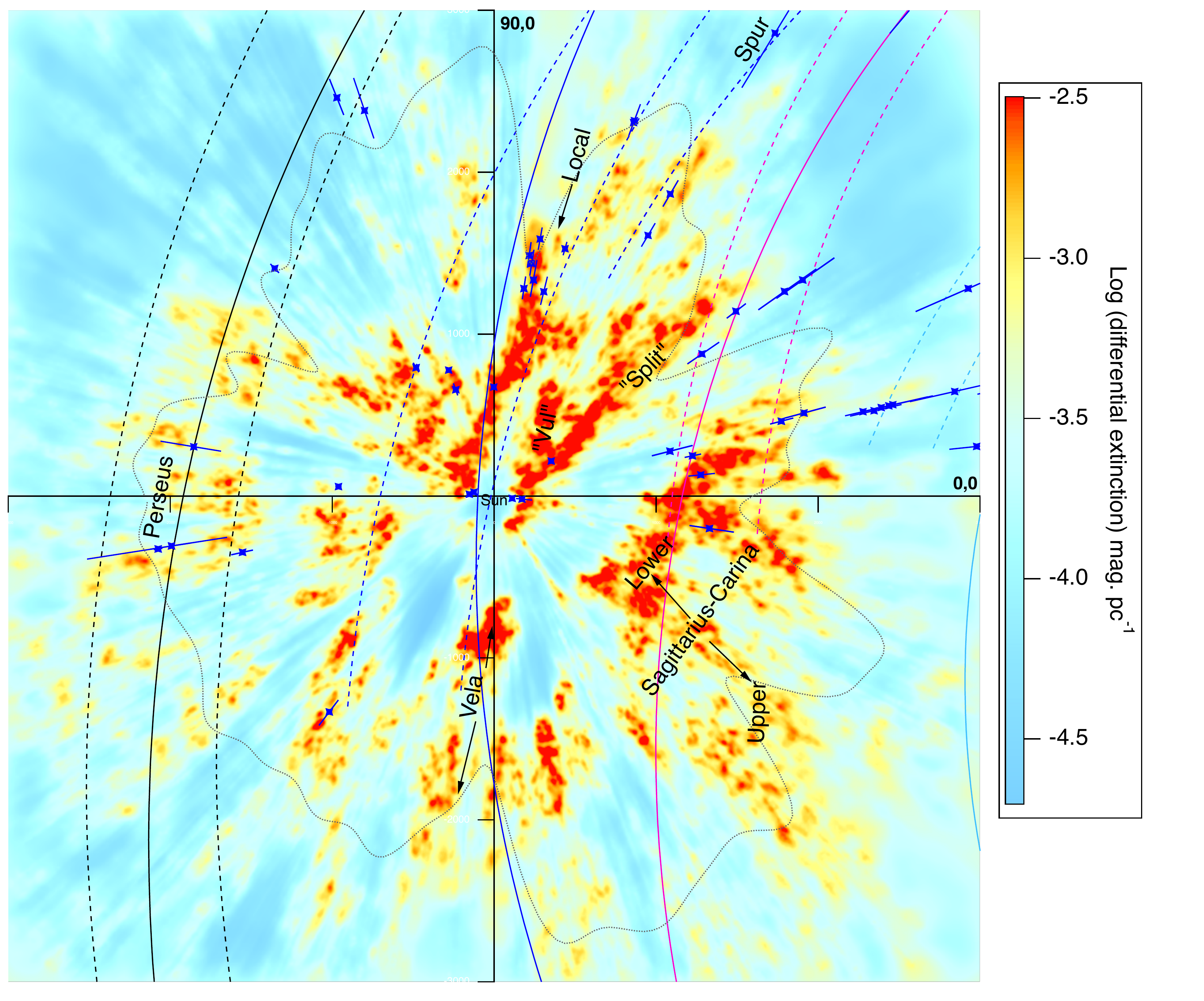}
\caption{Projections onto the Plane of the masers from \cite{Reid14}, \cite{Xu16} and \cite{Honma12} that possess parallax distances and are located within $50$~pc from the Plane, superimposed on the dust distribution along the Galactic plane. The Sun is at the center of the figure at (0,0). The Galactic center is to the right. The dotted gray line represents the distance beyond which the final resolution of 25 pc is not achievable due to target scarcity (see text). Shown are the spiral arms derived by \cite{Reid14} (dashed lines, inner and outer parts, together with the local spur of \cite{Xu16}) and \cite{HouHan14} (solid lines): black: Perseus, blue: Local Arm, magenta: Sagittarius, cyan: Scutum.}
\label{fig:masers_arms}
\end{figure}

 Looking at the solar plane in Figures \ref{fig:galplane} and \ref{fig:masers_arms}, a first prominent region is the well delineated inner part of the Carina-Sagittarius arm in the fourth quadrant and half of the first quadrant. Its series of compact dust cloud complexes is very well aligned along a l=45-225$\fdeg$ direction, i.e. a direction significantly tilted from the expected spiral arm direction. Such an orientation is already partly visible in the maps of \cite{Kos14} and \cite{Chen18}. In the fourth quadrant the clouds in front of this huge complex are as close as 1~kpc from the Sun. A more distant ($\sim$2kpc and more) and less compact dense dust area is also found beyond the first \emph{rank} of clouds, from l=300  to l=0$\fdeg$. The more diffuse aspect is likely partially due to the fact that this region is close to the limits of our maps and is mapped at lower resolution. There is no apparent, clear continuity between the two regions in the Plane. However, looking at the cloud distribution below and above the Plane is very informative. As a matter of fact, as shown by Figs. \ref{fig:south} and \ref{fig:north}, this second rank of structures is particularly prominent well above the solar plane and is oriented in a homogeneous way in the Z=+100 pc and even Z=+150 pc images, but its orientation is drastically different from the one of the large structure close to the Plane, here along a -90,+90$\fdeg$ axis. We have labeled it \emph{upper Sagittarius-Carina} in Fig. \ref{fig:masers_arms}.  The closest cloud complex is not apparent above the solar plane, however it is clearly seen below the Plane down to -150 pc. We labeled it \emph{lower Sagittarius-Carina}. Another way to look at this complex structure of dust clouds in Carina-Sagittarius is by inspection of the vertical planes of Fig. \ref{fig:vertical2}. The fourth quadrant corresponds to the left parts of the images. It  can be seen, especially along l=270 and l=230$\fdeg$ (left parts of first and seventh images) that there is a complex, wavy-type structure around the solar plane that is the counterpart of what is seen in horizontal planes. 
   
\begin{figure}[t]
\centering
\includegraphics[width=9cm]{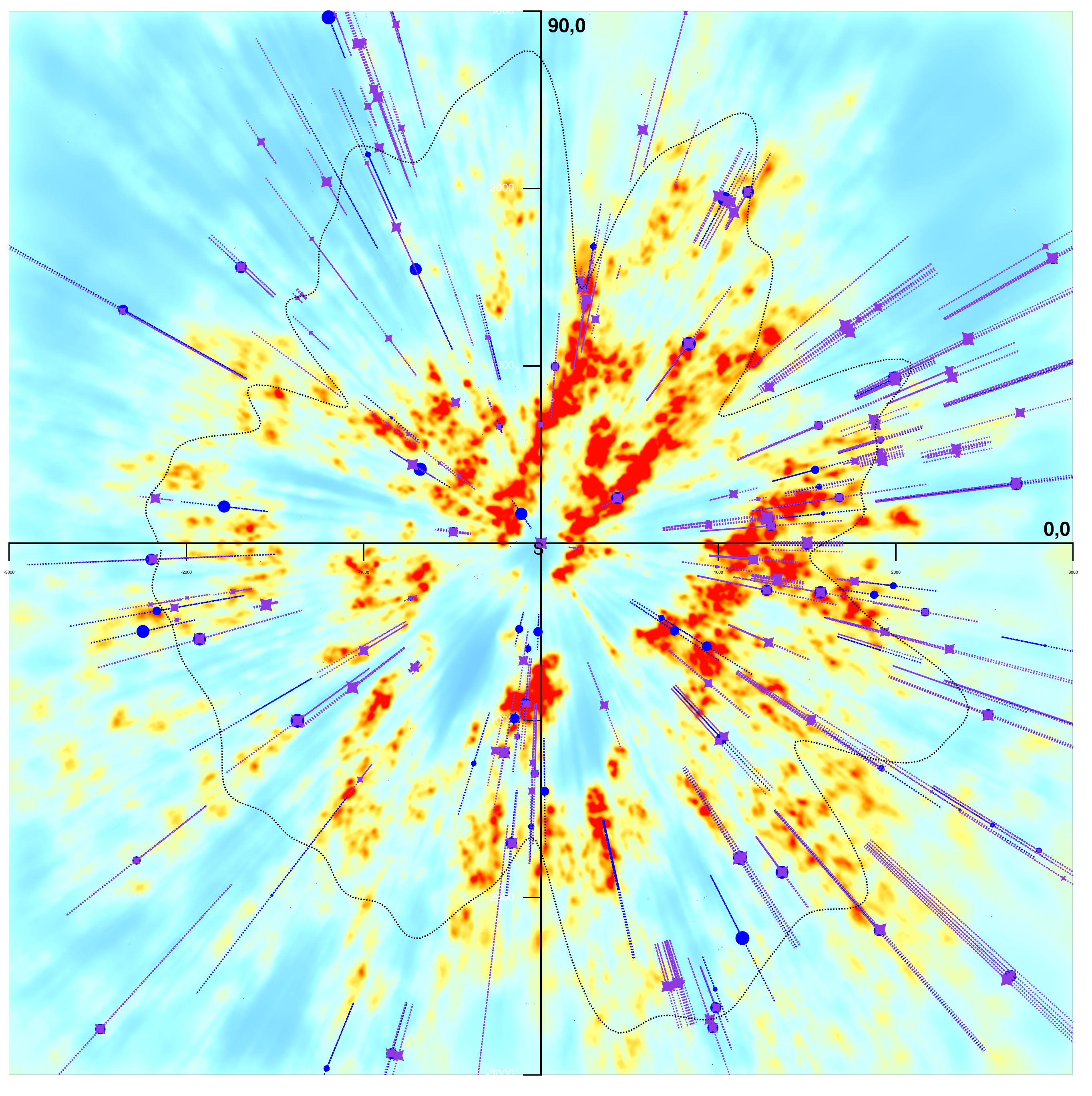}
\caption{Same as Fig. \ref{fig:masers_arms} for star forming regions from the compilation of \cite{HouHan14} and from \cite{Russeil03}.  Only objects with stellar distances (see text) and $|z|<50$~pc are shown and the size of the markers decreases for increasing $|z|$.}
\label{fig:planecomp_regions}
\end{figure}

\begin{figure}[t]
\centering
\includegraphics[width=9cm]{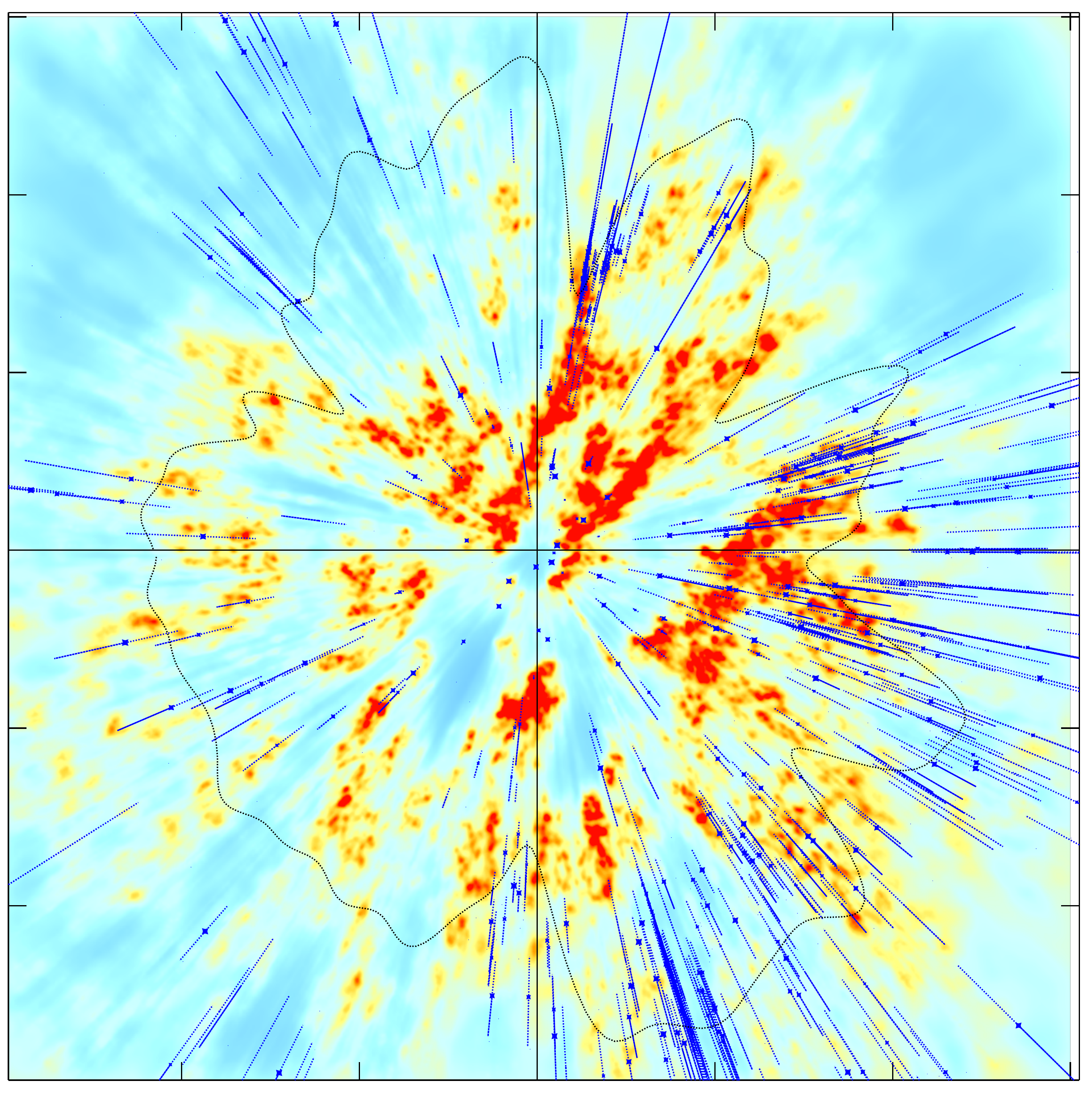}
\caption{Same as Fig. \ref{fig:masers_arms}  for O stars from the recent work of \cite{xu_comparison_2018}. Distances are from Gaia DR2. Only stars within 50 pc from the Plane are shown.}
\label{fig:xu_OBstars}
\end{figure}

\begin{figure}[t]
\centering
\includegraphics[width=9cm]{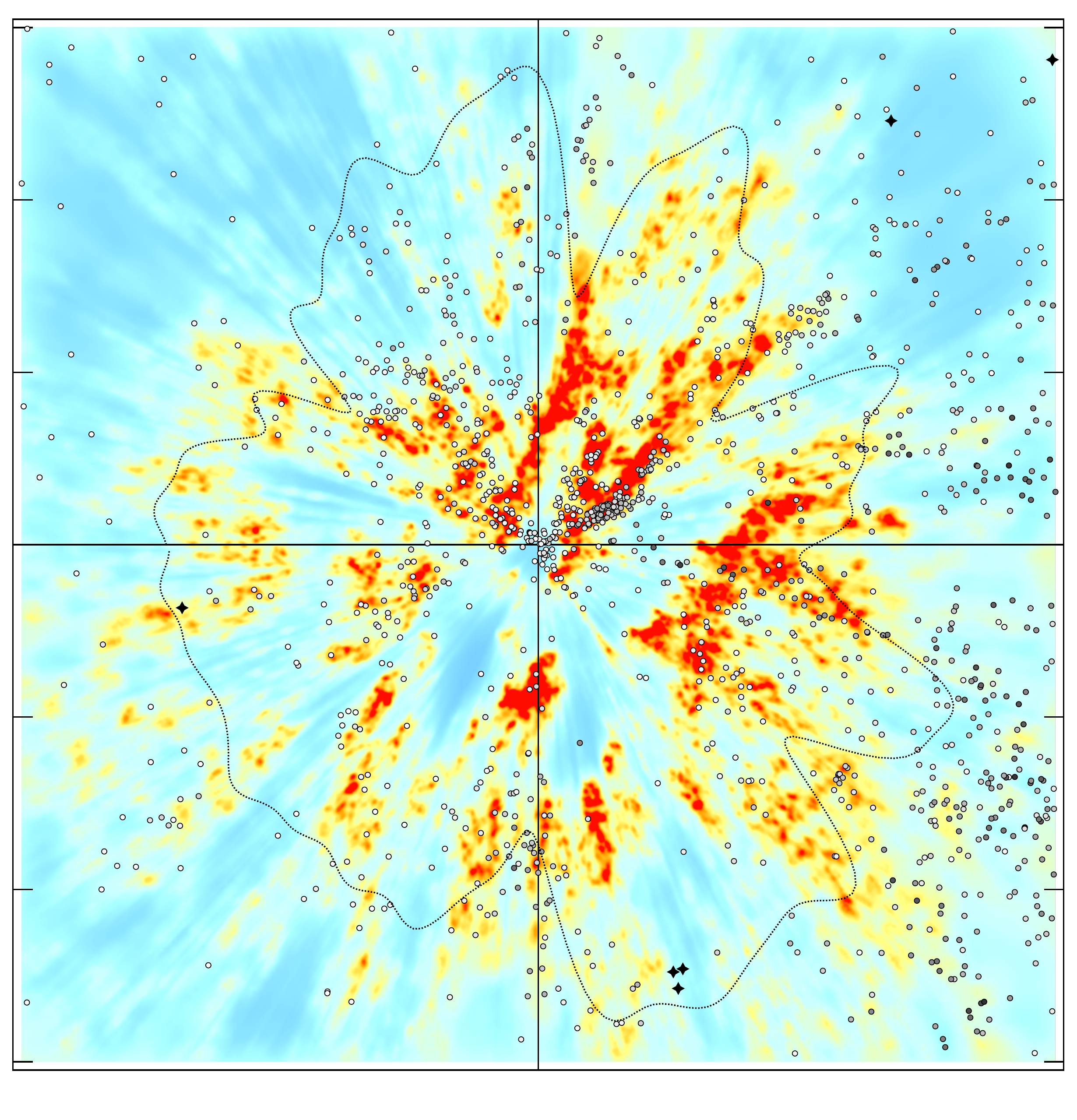}
\caption{Same as Fig. \ref{fig:masers_arms} for the CO clouds from \cite{miville-deschenes_physical_2016} (circles) and the nearby giant molecular clouds listed by \cite{HouHan14} (stars). The black and white scale indicates the CO equivalent width. Only objects with $|z|<50$~pc are shown and the size of the markers decreases for increasing $|z|$.}
\label{fig:MCs}
\end{figure}

   The second striking and prominent feature is the more than 2 kpc-long series of dense dust clouds towards l$\sim$45$\fdeg$. This structure that is close to, but not exactly radial and we nicknamed the \emph{split} is located between the Local Arm and the Carina-Sagittarius Arm. This structure is apparent at various degrees in the previous maps of \cite{Marshall06,Chen18,Sale14,Kos14,Green15,Green18}, but has never been commented. Interestingly, it has characteristics that differentiate it from other main regions. A look at Figures \ref{fig:planecomp_regions}, \ref{fig:xu_OBstars} and \ref{fig:masers_arms} shows almost no match between the \emph{split} and HII regions, masers or O star concentrations. On the contrary, Fig \ref{fig:MCs} reveals a very large concentration of molecular clouds all along the \emph{split}. Interestingly,  \cite{skowron_new_2018} also find a strong concentration of Cepheids right along the \emph{split}, that they attribute to a $\sim$30 million years old star formation burst. 
   
   A third, at least 2kpc long series of compact regions starts close to the Sun and is oriented along l=75-90 $\fdeg$. It corresponds to the well known Cygnus Rift and can be seen as the main part of the Local arm. Here again, the orientation of the structure varies with the distance Z to the mid-plane, as shown in Figures \ref{fig:south} and \ref{fig:north}. Below the mid-plane the orientation is towards l=$\sim$55 $\fdeg$, while at Z$\geq$50 pc it rotates and reach  l=$\sim$90 $\fdeg$.
   Interestingly, only the $\geq$ 1.2 kpc part of this structure is rich in O stars, HII regions, masers, and molecular clouds, while the closer part shows none of these tracers. 
   
   A shorter, about 0.8 kpc long structure is also apparent in the first quadrant, a structure anchored at the inner part of the 
   \emph{split} and roughly parallel to the Cygnus Rift, labeled \emph{Vul} in Fig \ref{fig:masers_arms}. Similarly to the \emph{split}, it is rich in MCs but tracers of star formation are absent. Farther away at d$\geq$1.5 kpc and towards $l\sim60\fdeg$ are the Vul OB4 masers and another elongated structure that could be the foreground counterpart of the \emph{local spur} of \cite{Xu16}.

 In the second  and third quadrants there are series of clouds that are approximately  aligned along the rotation direction and located at about 2 kpc distance from the l=90-270 axis. These complexes correspond to the expected location of  the Perseus arm. They are distributed in a much less compact way than the structures previously described, suggesting that the outer Perseus is a more diffuse arm. Surprisingly, only very few dust clouds are seen between l=120$\fdeg$ and l=150$\fdeg$ where one would expect a  prolongation of Perseus and clouds at about 2 kpc. Indeed, the existence of the Perseus arm is indicated in this region  by several HII regions, O stars and masers. A potential explanation could be a bias of our mapping due to scarcity of targets in this area, due to the high opacity of clouds in the foreground.  However the threshold white line in Fig. \ref{fig:galplane} shows that other regions with the same foreground opacity have well defined dust clouds. Moreover, there are no detected MCs in this area, suggesting that both dust and molecular gas have been processed due to strong star formation.

The inner part of the third quadrant is characterized by the giant cavity centered on l=240$\fdeg$, already clearly defined in our previous 3D maps. This cavity is surrounded by the main star forming regions of Orion and Vela. It is adjacent and connected to the Orion-Eridanus cavity below the Plane. The latter is better seen in the third vertical plane in Fig \ref{fig:vertical1}. At all distances from the Plane this region of the third quadrant extends between 150 pc to 1.2 kpc. Note that this region is devoid of dust but is filled with ionized gas \citep{Lall47tuc}, suggesting that most of the dust evaporated during the event that ionized the gas. As we already noticed in previous maps, this cavity is aligned with a narrower cavity elongated along l=60$\fdeg$, i.e. a cavity separating the Cygnus Rift and the \emph{split}. This axis also corresponds to an ionization gradient in the nearby ISM \citep{Wolff99}.

\section{Discussion and perspectives}\label{Discussion}

We have used the Gaia and 2MASS photometric data to estimate the extinction towards $\sim$27 millions carefully selected target stars having a Gaia DR2 parallax uncertainty below 20\%. $\sim$16 millions of distance-extinction pairs for the closest objects have been inverted to produce 3D maps of the dust density within 400 pc from the Plane and in a 6 kpc by 6 kpc region along the Plane. To do so, a new hierarchical  inversion algorithm  has been developed that has the advantage of adapting the map resolution to the target space density. Following this procedure, the minimum size of the inverted structures varies from 25 pc (a resolution achieved within 1 kpc) to 500 pc (in a few regions distant by more than 3kpc from the Sun).  Our online tools  (\url{http://stilism.obspm.fr}) will be updated to include the new 3D map. They allow one to draw an image of the dust distribution in any chosen plane, containing  the Sun or not, as well as reddening profiles.  Images and  cumulative extinctions as well as the 3D distribution itself will be available for download. We emphasize that such full 3D inversions have the advantage of showing how cloud complexes are related to each other in 3D space. As we already warned in previous similar works, the absolute values of the integrated extinction between the Sun and each point in the computation volume may underestimate the extinction in the cases of sightlines crossing cloud cores. This loss of angular resolution is due to the regularization constraints and is the price to pay for using full 3D correlations between neighboring points. \textbf{Moreover, due to the general bias toward less obscured, hence brighter stars, the differential extinction is likely to be biased low, especially at large distances.}

 The dust density distribution appears dominated by elongated structures with a double system of orientations. The first quadrant is dominated by two elongated structures. One of the two is coincident with the well known Cygnus Rift, usually identified as the local arm. Its orientation varies from close to the direction of rotation l=80-90$\fdeg$ above the mid-plane to about l=60$\fdeg$ below the mid-plane. The second structure, here nicknamed the \emph{split}, is oriented very differently from the direction of rotation and is seen as a long arm segment linking the Perseus and Sagittarius-Carina spiral arms \citep[see e.g.][]{HouHan14, vallee_guided_2017}. The latter structure is different from the \emph{spur} seen by \cite{Xu16} that is mostly further away from our map and may be seen as the prolongation of a third group of clouds starting beyond 1.5 kpc (see Fig. \ref{fig:masers_arms}). The local arm and the \emph{split} are both strong in extinction, but the local arm is better traced with masers and star formation regions, especially at large distance  \citep{Reid14}, while the \emph{split} is stronger in CO \citep{miville-deschenes_physical_2016}. According to the Cepheids study of \cite{skowron_new_2018} there seems to be an age difference between those structures, the split being younger.

 The new map reveals a particularly compact and well delineated foreground part of Sagittarius-Carina that extends in the fourth quadrant and at $0<l<30\fdeg$ in the first quadrant. This structure labeled \emph{inner} in Fig. \ref{fig:masers_arms} is mainly seen within and below the Plane and has the same orientation as the \emph{split}. A second and similarly compact \emph{outer} part of Sagittarius-Carina is located at larger distance and predominantly 50 to 150 pc above the Plane. At  variance with the inner part it is oriented along the direction of rotation. 

 A large fraction of the Perseus arm is perceptible in the second and third quadrants, but it appears significantly more fragmented than the other arm segments. In the second quadrant it is not detected over for $120<l<145\fdeg$, in agreement with the small number of molecular clouds in this area but in contrast with HII regions and masers. Whether the absence of dust and CO is related to past star formation or there is a bias in dust detection deserves further study.
 
 The distribution of the dust along directions perpendicular to the Plane varies strongly from one region to the other, and there are striking structures, having in some regions a wavy pattern with periods on the order of 1-2 kpc, as revealed by images in vertical planes (Fig \ref{fig:vertical1} and \ref{fig:vertical2}). The largest amplitude of off-plane excursions is found in the Sun vicinity, within about 500 pc. It can hardly be an observing bias  related to obscuring foreground clouds. However, additional data will be necessary to preclude biases linked to scarcity of targets at large distance from the mid-plane. Interestingly, stellar velocity patterns are also particularly complex within 500 pc (see e.g. \cite{Kawata18}, \cite{Antoja18}, \cite{GaiacollKatz18}). Finally, many more cloud structures and cavities devoid of dust are revealed by the 3D inversion but their description is beyond the scope of this paper. 

 In the future, improved parallax measurements and more numerous extinction estimates are expected, allowing  inversions at  better spatial resolution and the extension of the maps to larger distances. For such maps, it will become necessary to include estimates of bias in parallax distance at the individual level, and this is part of our current efforts. The models of Galactic evolution, currently being revolutionized by Gaia, will certainly benefit from detailed comparisons between the distributions and motions of stars and interstellar matter. 3D dust maps such as our inverted maps can be useful directly, or indirectly as ingredients for new kinetic tomography techniques for the gaseous ISM \citep{Tchernyshyov17} that use 3D cloud distributions and CO/HI radio spectral cubes and/or absorption data to assign velocities to interstellar clouds. This in turn will allow detailed comparisons between the stellar and ISM motions and more constraints on star formation.

\begin{acknowledgements}
This work has made use of data from the European Space Agency (ESA) mission Gaia (https://www.cosmos.esa.int/gaia), processed by the Gaia Data Processing and Analysis Consortium (DPAC, https://www.cosmos.esa.int/web/gaia/dpac/consortium). Funding for the DPAC has been provided by national institutions, in particular the institutions participating in the Gaia Multilateral
Agreement. This work also makes use of
data products from the 2MASS, which is a joint project of
the University of Massachusetts and the Infrared Processing
and Analysis Center/California Institute of Technology,
funded by the National Aeronautics and Space Administration
and the National Science Foundation.
\end{acknowledgements}
\bibliographystyle{aa}
\bibliography{mybib.bib}




\appendix
\section{Reddening profiles and comparisons with profiles 
from Pan-STARRS~1 and 2MASS}
\begin{figure*}[t]
\includegraphics[width=18.7cm]{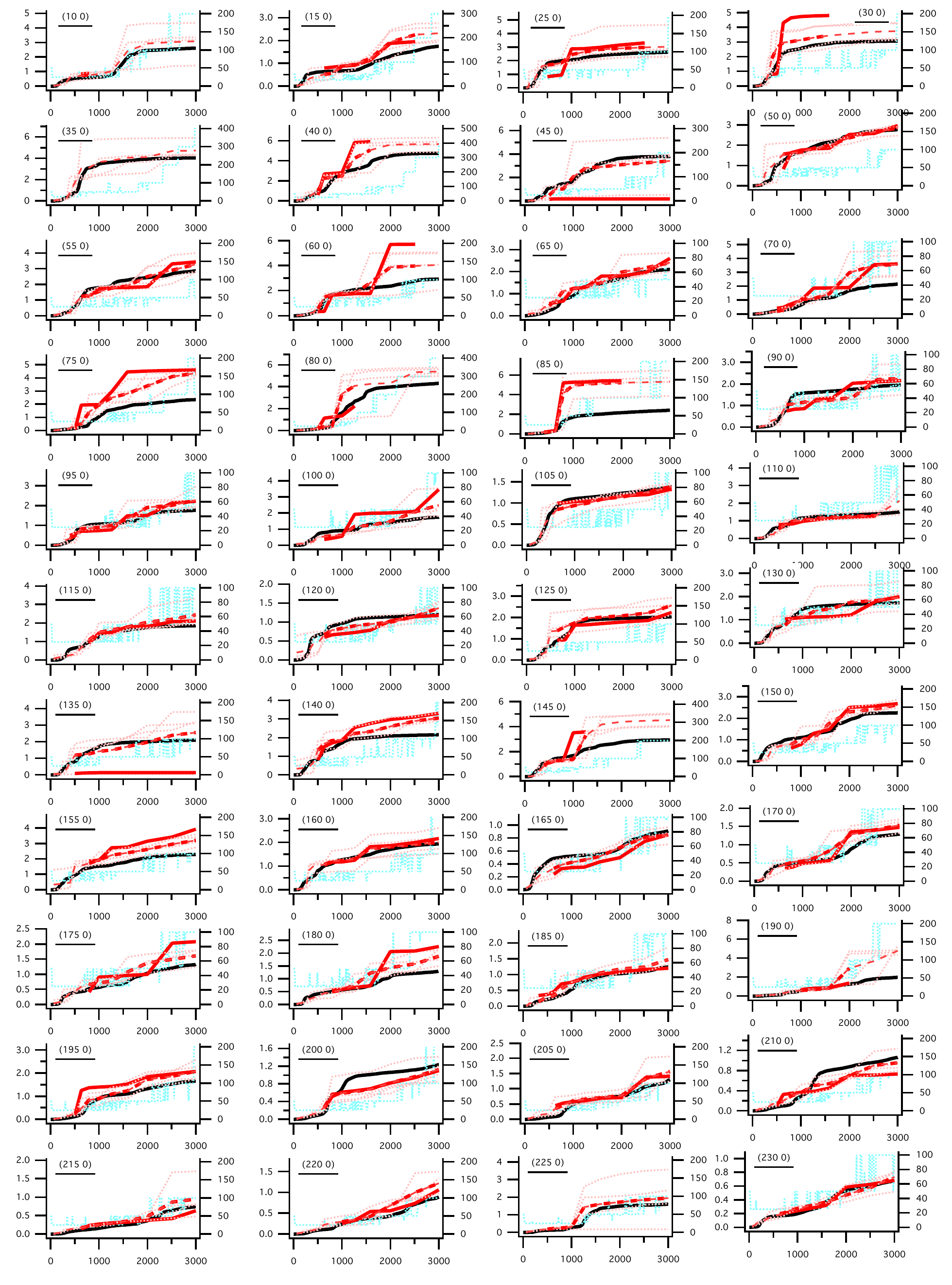}
\caption{Extinction profiles integrated through the 3D distribution of differential extinction for sightlines within the Plane (black) and reddening profiles from \cite{Green18} scaled to A$_{0}$ (see text, red solid). Also shown are the \cite{Green18} profiles for 4 surrounding directions at $\pm$ 0.25 $\fdeg$ in longitude and latitude (dotted red)  and their average in non-flagged and all intervals resp. (thick (resp. thin dashed red). Our achieved resolution (see section \ref{inversions}) is also shown (pale blue, right scale. Coordinates are indicated on the graph.}
\label{fig:comppanstarrsnew}
\end{figure*}

\cite{Green18} have made available reddening profiles derived from Pan-STARRS~1 (PS1) and 2MASS for directions spaced by about 10 arcmn in the whole sky region accessible to Pan-STARRS. It is particularly interesting to compare 3D maps based on different photometric systems and different techniques. In order to perform comparisons between our results based on Gaia $G$,$\BP$,$\RP$, and 2MASS and the \cite{Green18} reddening profiles we have integrated our computed differential reddening along a series of directions. We have converted uniformly the color excess E(B-V) from the \cite{Green18} map into A$_{0}$ by multiplying by a factor 0.88*3.1, i.e. assuming R=3.1 and following the scaling factors recommended by the authors.  \textbf{For all the comparisons we used their maximum-probability density estimate (the best-fit) distance-reddening curve}. Fig. \ref{fig:comppanstarrsnew} displays comparisons within the Plane every 5\fdeg\ in the longitude interval l=10-230\fdeg. In addition to the \cite{Green18} central profile at the exact coordinates (l,b) we have added their profiles for longitudes l$\pm$0.25\fdeg\ and b$\pm$0.25\fdeg\ as well as the average between the five profiles. Our maximum resolution is also indicated as a function of distance on the right scale. Both the adjacent directions and the resolution allow to better interpret the comparisons. The figure shows that at distances where our achieved resolution is 25~pc or 50~pc our reddening profile compares well with either the central PS1 profile or the average profile. Marked differences appear where our achieved resolution is poorer (from 100~pc) and in very complex regions with a large number of intervening clouds, e.g. between longitudes l=75\fdeg\ and 85\fdeg\ beyond 500-1000~pc. In those case clouds are similarly located, however our extinction is underestimated. On the other hand, the good agreement observed at short distances where the PS1 flag starts to be OFF suggests that our measurements within several hundreds pc are in good continuity with PS1.


%

\end{document}